\documentclass[sigconf]{acmart}

\AtBeginDocument{%
  }

\copyrightyear{2025}
\acmYear{2025}
\setcopyright{cc}
\setcctype{by}
\acmConference[CHI '25]{CHI Conference on Human Factors in Computing
Systems}{April 26-May 01, 2025}{Yokohama, Japan}
\acmBooktitle{CHI Conference on Human Factors in Computing Systems (CHI
'25), April 26-May 01, 2025, Yokohama,
Japan}
\acmDOI{10.1145/3706598.3713953}
\acmISBN{979-8-4007-1394-1/25/04}

\usepackage{enumitem}

\newif\ifcomment
\commentfalse

\ifcomment
\newcommand{\hide}[1]{}
\newcommand{\note}[1]{\textcolor{blue}{<< #1 >>}}
\newcommand{\added}[1]{\textcolor[rgb]{0.1, 0.56, 1}{#1}}

\newcommand{\cut}[1]{\textcolor[rgb]{0.5,0.5,0.5}{CUT: #1}}

\newcommand{\todo}[1]{\textcolor{red}{TODO: #1}}

\newcommand{\deleted}[1]{\textcolor[rgb]{0.8,0.8,0.8}{#1}}

\newcommand{\nuwan}[1]{\textcolor{blue}{NUWAN: #1}}
\newcommand{\runze}[1]{\textcolor[rgb]{0.5,0.2,0.2}{RUNZE: #1}}
\newcommand{\zsd}[1]{\textcolor[rgb]{0.5,0.1,0.8}{ZSD: #1}}
\newcommand{\danny}[1]{\textcolor[rgb]{0,0.8,0.4}{DANNY: #1}}

\else
\newcommand{\hide}[1]{}
\newcommand{\note}[1]{}
\newcommand{\added}[1]{#1}

\newcommand{\cut}[1]{}

\newcommand{\todo}[1]{}

\newcommand{\deleted}[1]{}

\newcommand{\nuwan}[1]{}
\newcommand{\runze}[1]{}
\newcommand{\zsd}[1]{}
\newcommand{\danny}[1]{}

\fi

\newcommand{\meansd}[2]{$M = #1,\: SD = #2$}

\newcommand{\pval}[1]{$p#1$}

\renewcommand{\quote}[1]{``#1''}
\newcommand{\quoteby}[2]{``#2 (#1)''}

\newcommand{\significantI}[0]{$^{\star}$}

\newcommand{\PhaseI}[0]{\textit{Phase 1}}
\newcommand{\PhaseII}[0]{\textit{Phase 2}}

\newcommand{\AiGet}[0]{\textit{AiGet}}
\newcommand{\BaselinewoR}[0]{\textit{Baseline w/o R}}
\newcommand{\BaselinewoRP}[0]{\textit{Baseline w/o RP}}

\newcommand{\assistant}[1]{\textit{assistant}{#1}}

\newcommand{\PrioritizationFormula}[0]{$\textbf{Novelty}$ $\times$ ($\textbf{AlignUserPreference}$ + $\textbf{Utility}$ + $\textbf{Unexpectedness}$)}

\newcommand{\enjoyment}[0]{\textit{Enjoyment of Primary Task}}
\newcommand{\distraction}[0]{\textit{Distraction}}
\newcommand{\naturalness}[0]{\textit{Naturalness}}

\newcommand{\overallScore}[0]{\textit{Overall Perceived Scores}}
\newcommand{\novelty}[0]{\textit{Novelty}}
\newcommand{\personalization}[0]{\textit{Personalization}}
\newcommand{\relevance}[0]{\textit{Relevance}}
\newcommand{\unexpectedness}[0]{\textit{Unexpectedness}}
\newcommand{\usefulness}[0]{\textit{Usefulness}}
\newcommand{\notAnnoying}[0]{\textit{Not Annoying}}
\newcommand{\notOverload}[0]{\textit{Not Overload}}

\newcommand{\interesting}[0]{\textit{Interesting}}
\newcommand{\deepenKnowledge}[0]{\textit{Deepen Topic Understanding}}
\newcommand{\deepenEnvironment}[0]{\textit{Increased Environment Connection}}

\newcommand{\SUS}[0]{\textit{SUS}}
\newcommand{\perceivedTaskLoad}[0]{\textit{RTLX}}

\definecolor{mypurple}{RGB}{128, 0, 255} %

\newcommand{\saccade}[1]{\textit{Saccade}{#1}}
\newcommand{\quickBrowse}[1]{\textit{Quick Browse}{#1}}
\newcommand{\focused}[1]{\textit{Focused}{#1}}

\newcommand{\designGoalContextUnderstanding}[0]{\hyperref[sec:study1:design_goals]{\textbf{\textit{D1}}}}
\newcommand{\designGoalPersonalization}[0]{\hyperref[sec:study1:design_goals]{\textbf{\textit{D2}}}}
\newcommand{\designGoalPrioritization}[0]{\hyperref[sec:study1:design_goals]{\textbf{\textit{D3}}}}
\newcommand{\designGoalMinimizedDistraction}[0]{\hyperref[sec:study1:design_goals]{\textbf{\textit{D4}}}}

\begin{document}

\title[\textit{AiGet}]{\AiGet{}: Transforming Everyday Moments into Hidden Knowledge Discovery with AI Assistance on Smart Glasses}

\author{Runze Cai}
\orcid{0000-0003-0974-3751}
\email{runze.cai@u.nus.edu}
\affiliation{%
  \institution{Synteraction Lab}
  \institution{School of Computing, National University of Singapore}
  \country{Singapore}}

\author{Nuwan Janaka}

\orcid{0000-0003-2983-6808}
\email{nuwanj@u.nus.edu}
\affiliation{%
  \institution{Synteraction Lab}
  \institution{Smart Systems Institute, National University of Singapore}
  \country{Singapore}
}

\author{Hyeongcheol Kim}
\authornote{Both authors contributed equally to this research.}
\orcid{0000-0003-4327-2148}
\email{hyeongcheol@u.nus.edu}
\affiliation{%
\institution{Synteraction Lab}
  \institution{National University of Singapore}
  \country{Singapore}
}

\author{Yang Chen}
\authornotemark[1]
\orcid{0000-0003-3129-8447}
\email{cyang@u.nus.edu}
\affiliation{%
\institution{College of Design and Engineering, National University of Singapore}
  \country{Singapore}
}

\author{Shengdong Zhao}
\authornote{Corresponding Authors.}
\orcid{0000-0001-7971-3107}
\email{shengdong.zhao@cityu.edu.hk}
\affiliation{%
\institution{Synteraction Lab}
\institution{School of Creative Media \& Department of Computer Science, City University of Hong Kong}
\city{Hong Kong}
  \country{China}
}

\author{Yun Huang}
\orcid{0000-0003-0399-8032}
\email{yunhuang@illinois.edu}
\affiliation{%
\institution{School of Information Sciences, University of Illinois at Urbana-Champaign}
\city{Champaign}
\state{Illinois}
  \country{United States}
}

\author{David Hsu}
\authornotemark[2]
\orcid{0000-0002-2309-4535}
\email{dyhsu@comp.nus.edu.sg}
\affiliation{%
\institution{School of Computing, Smart Systems Institute, National University of Singapore}
\country{Singapore}
}
\renewcommand{\shortauthors}{Cai et al.}

\begin{abstract}
Unlike the free exploration of childhood, the demands of daily life reduce our motivation to explore our surroundings, leading to missed opportunities for informal learning. Traditional tools for knowledge acquisition are reactive, relying on user initiative and limiting their ability to uncover hidden interests. Through formative studies, we introduce \AiGet{}, a proactive AI assistant integrated with AR smart glasses, designed to seamlessly embed informal learning into low-demand daily activities (e.g., casual walking and shopping). \AiGet{} analyzes real-time user gaze patterns, environmental context, and user profiles, leveraging large language models to deliver personalized, context-aware knowledge with low disruption to primary tasks. In-lab evaluations and real-world testing, including continued use over multiple days, demonstrate \AiGet{}’s effectiveness in uncovering overlooked yet surprising interests, enhancing primary task enjoyment, reviving curiosity, and deepening connections with the environment. We further propose design guidelines for AI-assisted informal learning, focused on transforming everyday moments into enriching learning experiences.

\end{abstract}



\maketitle

\begin{figure*}
  \includegraphics[width=\textwidth]{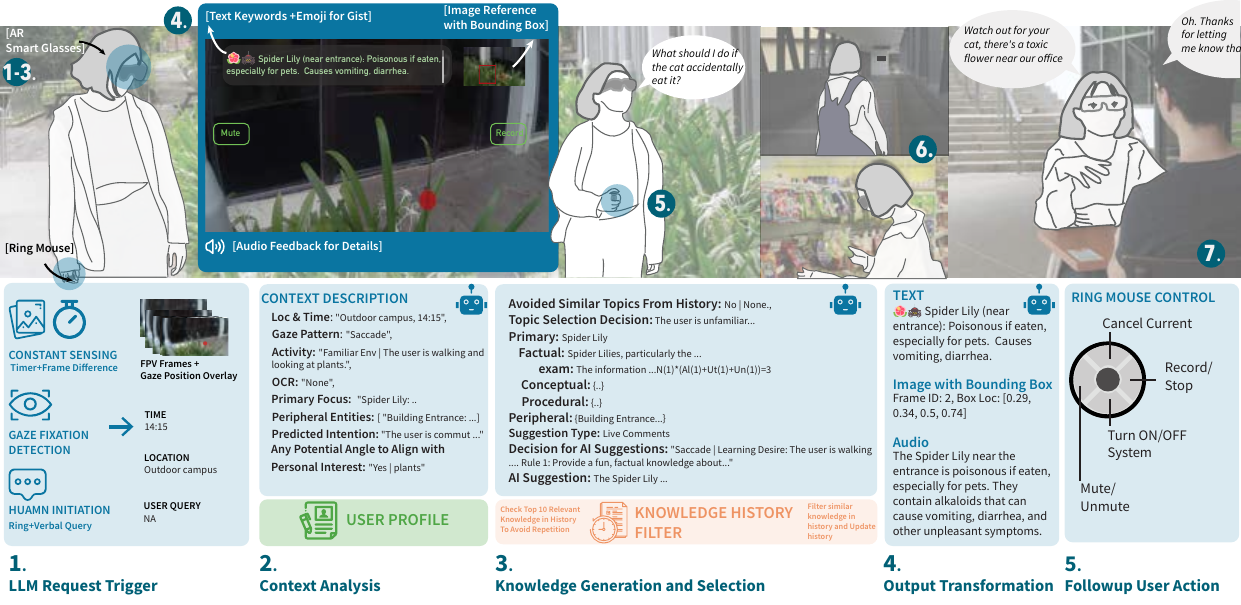}
  \caption{A day with \AiGet{}, a wearable knowledge discovery assistant equipped with AR smart glasses and a ring mouse. (1) While casually walking, \AiGet{} actively analyzes the user's gaze patterns and identifies ``interesting'' environmental entities. (2) \AiGet{} recognizes the spider lily that the user briefly glanced at as a valuable opportunity for informal learning, considering her interest in plants in the User Profile and lack of knowledge about the flower. (3) Using contextual data, the system generates multiple knowledge candidates and selects the most valuable one about the spider lily for the user. (4) \AiGet{} delivers multimodal feedback, including audio, text keywords with emojis, and an image with a bounding box for the flower to help the user locate the flower and gain information with minimal interference to primary tasks. (5) Surprised by the knowledge delivery about the overlooked flower, the user uses the ring mouse to ask follow-up questions. (6) Later, \AiGet{} continuously helps her uncover design details behind the campus architecture and compare two unfamiliar snacks, aligning with her interest in healthy diets, etc. (7) When the user meets a friend who owns a cat, she shares what she learned from \AiGet{}---that spider lilies are toxic to pets---enriching their social interaction.}
  \Description{This figure depicts a user interaction scenario with AiGet, a wearable knowledge discovery assistant that combines AR smart glasses and a ring mouse for interaction. The user, while walking outdoors, briefly glances at a Spider Lily plant. AiGet, using gaze detection, identifies this as an opportunity for contextual learning based on the user’s interest in plants and lack of prior knowledge about the flower. The system processes contextual information, including time, location, gaze patterns, and past interactions, to generate knowledge suggestions. AiGet then selects the most relevant piece of information—highlighting the plant’s toxicity to pets—and delivers it through a combination of audio feedback, visual text, and an AR bounding box to identify the plant. The user, intrigued by the information, engages with the system using the ring mouse to ask follow-up questions. Later, AiGet assists the user with other tasks, such as discovering architectural details on campus and comparing unfamiliar snacks. Finally, the user shares the knowledge about the plant's toxicity with a friend who owns a cat, enriching their social interaction. This scenario showcases AiGet’s ability to seamlessly integrate contextual learning into everyday experiences.}
  \label{fig:teaser}
\end{figure*}

\section{Introduction}

\textit{\quote{The real voyage of discovery consists not in seeking new landscapes, but in having new eyes.} – Marcel Proust}.

Our daily surroundings are filled with rich knowledge, from building designs to local flora and fauna, offering countless opportunities for discovery and informal learning from everyday moments \cite{kaplan1989experience_nature,bekerman2006learning_in_places, ardoin2021environmental}. Such informal learning not only allows us to gather useful knowledge about our surroundings but also fosters critical thinking and diverse perspectives, helping us adapt to new situations and inspire innovation \cite{marsick_informal_2001, cross_informal_2007, eraut__informal_2004}. Yet, modern adult life often disconnects us from these rich opportunities, making us ``blind'' to such discoveries. Pressed by daily demands, adults focus on primary tasks with higher priority (e.g., rushing for work and studies), bypassing opportunities for curiosity and exploration. We also tend to prioritize immediate rewards, avoiding the perceived effort of learning \cite{liquin2022exploitation}. Moreover, the rise of smartphones and social media, which offer effortless access to knowledge, has further diverted our attention from the physical world. As a result, we may overlook rare migratory birds in our backyard while being well-versed in exotic species from online videos; we may enjoy cooking videos about exotic ingredients but remain unaware of unique vegetables like purple broccoli available at the local supermarket. 

Emerging technologies like Multimodal Large Language Models (MLLMs) \cite{yin2023survey} and wearable devices, such as smart glasses, offer new possibilities for "having new eyes" to discover hidden knowledge in our environment. By using cameras on smart glasses to sense the surroundings, MLLMs, with their vast knowledge base, can identify relevant information about entities in the environment and proactively deliver it via smart glasses, even when users are engaged in other tasks like commuting. However, this approach faces two challenges: (1) given the multitude of elements that can be explored in our surrounding environment, it remains unclear which types of knowledge users value most in everyday moments; (2) although proactive knowledge delivery may reduce the effort for manual searches (e.g., using Google or ChatGPT), it could also disrupt primary tasks if not carefully designed. These challenges may hinder the adoption of proactive wearable knowledge discovery assistants. As a result, existing work has largely focused either on user-initiated queries (e.g., users look at a target and ask for information) \cite{wang2024G-VOILA,lee2024gazepointAR, google_project_2024}, or limited learning content, such as foreign vocabulary \cite{Arakawa_Yakura_Kobayashi_2022} or pre-registered material (e.g., AR in museums~\cite{Sommerauer_Müller_2014}), limiting deeper engagement with the everyday environment.

This leads to our main research question and design goal: \textit{How can we utilize wearable intelligent assistants to help users rediscover the \textbf{hidden} yet \textbf{desired} knowledge embedded in the surrounding environment with \textbf{reduced interference to their primary tasks} in everyday settings?}

To answer this question, we first conducted a formative study to identify users' learning interests and challenges in everyday contexts. The results revealed four cognitive biases and limitations that hinder users from discovering hidden knowledge: time constraints, inattentional blindness, lack of motivation to explore familiar environments, and a preference for staying within comfort zones rather than exploring unfamiliar ones \cite{mack2003inattentional, baron2023thinking, money2003uncertainty_avoid, rozenblit2002illusion_explanatory_depth}. Despite these challenges, participants consistently expressed enthusiasm for a proactive wearable assistant capable of identifying ``unseen'' and ``unknown'' knowledge across different contexts.

Building on these insights, we designed and implemented \AiGet{}, a proactive AI assistant integrated with AR smart glasses, addressing these limitations by seamlessly embedding learning opportunities into daily activities. It analyzes real-time user gaze patterns alongside environmental context and fine-tunes knowledge using personalized user profiles. While existing systems often rely on user-initiated interactions or focus solely on objects in the user's direct line of sight, \AiGet{} extends these capabilities by proactively identifying useful insights about entities in the broader visual field, including areas not under the user's immediate attention. This approach accommodates both ``unseen'' and ``seen but unknown'' knowledge, enabling users to discover valuable information they might have otherwise missed. 

\AiGet{}'s design is carefully tailored for everyday mobile scenarios, ensuring seamless integration alongside primary tasks, with two key aspects:
1) \textit{Information Content}: To enhance perceived usefulness and reduce annoyance, \AiGet{} prioritizes knowledge that balances novelty, personalization, usefulness, and unexpectedness \cite{kelly_individual_2021, sharot2020utility_decision_making,kotkov2023serendipity}.
2) \textit{User Interface Design}: To reduce interference and maintain environmental awareness, the system employs multimodal output: keywords and emojis for gist, images with bounding boxes for spatial reference, and audio for details, making knowledge absorption easier.
To provide user control while minimizing conflicts during undesired learning moments (e.g., rushing through time-sensitive errands), \AiGet{} uses a ring mouse and subtle gestures for interactions such as turning the system on/off, canceling displays, or initiating follow-up queries.
This integrated design naturally embeds learning opportunities into daily activities while remaining unobtrusive and user-friendly.

We conducted in-lab simulations with 12 participants and real-world evaluations involving 18 participants (with 6 for continued usage for up to 7 days) in low-demand daily activities\footnote{In our context, we focused on daily tasks with relatively low temporal or cognitive pressure, excluding tasks such as focused work to avoid overwhelming users.}, such as casual walking, shopping, and museum exploration. \AiGet{} demonstrated effectiveness in uncovering overlooked knowledge in environments, providing information that expands users' knowledge bases while being useful, personalized, and surprising. As a result, it enhanced primary task enjoyment through relevant and timely information delivery, revived users' attention and curiosity about surroundings, deepened their connection with their environment, and demonstrated potential for long-term usability across diverse daily contexts.

Our contributions are threefold:
(1) Design insights for analyzing the real-time multimodal environmental and user context to facilitate informal learning in everyday settings, addressing common barriers to spontaneous knowledge acquisition.
(2) \AiGet{}, a proof-of-concept wearable intelligent assistant carefully designed for mobile everyday usage, transforming daily moments into knowledge discovery opportunities and supporting the acquisition of personalized, context-relevant information.
(3) Empirical insights into how AI-assisted knowledge discovery can enhance engagement with one's environment by identifying ``unseen and unknown'' knowledge, reviving curiosity, and increasing attention to surroundings, along with design implications for future wearable informal learning assistants.

\section{Related Work}
\label{sec:related_work}
This section provides: (1) an overview of learning in daily life and the computing technologies that support it, (2) recent trends in using AI to enhance learning in everyday contexts, and (3) wearable solutions that facilitate learning during daily activities.

\subsection{Computing Technologies for Learning in Daily Life}
\label{sec:related_work_conventional_tech_support}

Learning occurs not only in formal settings like schools, with instructors and structured curricula, but also \textbf{informally} in daily life, where knowledge and skills are gained through everyday experiences and interactions \cite{jarvis2009learning, conlon2004review}. 
Such informal learning can occur either \textbf{intentionally} (a.k.a. self-directed, where learners initiate their learning) or \textbf{incidentally} (a.k.a. serendipitous, where learning occurs unexpectedly or unintentionally, often as a result of chance encounters), at any time and place within daily contexts~\cite{conlon2004review}. It, thereby, fosters more flexible learning in everyday environments \cite{marsick_informal_2001, ardoin2021environmental}. However, such informal learning poses challenges such as maintaining motivation and engagement \cite{song_motivational_2016} and balancing learning with the time constraints of daily activities \cite{marsick_informal_2001, livingstone_adults_2001}.

Ubiquitous computing devices, including mobile phones and Augmented/Mixed Reality (AR/MR) head-mounted displays (HMDs), have enabled learning at any time and place \cite{Yahya_Ahmad_Jalil_2010}, thereby increasing computer-assisted learning opportunities \cite{Attewell_Savill-Smith_2004, Lee_2023}. The sensing capabilities of these devices also allow for contextual awareness, enhancing both intentional and incidental daily learning by aligning content with learners' contexts \cite{edge_micromandarin_2011, cai_waitsuite_2017, hautasaari_vocabura_2020}.
Supportive systems that leverage context-awareness for daily learning include identifying suitable learning moments \cite{kim_interrupting_2024, janaka_visual_2022, cai_waitsuite_2017}, delivering context-specific learning content \cite{edge_micromandarin_2011, hutchison_context-sensitive_2007, Draxler_Labrie_Schmidt_Chuang_2020, Sommerauer_Müller_2014, Vazquez_Nyati_Luh_Fu_Aikawa_Maes_2017}, and providing manageable, situational knowledge \cite{Lee_2023}. These approaches have been shown to enhance learners' motivation, engagement, and knowledge retention \cite{sawyer_cambridge_2022, mayer_cambridge_2014}.

Despite these advances, traditional informal learning systems are often topic- or domain-specific, such as foreign languages \cite{hautasaari_vocabura_2020}, courses \cite{brahimi2015learning}, or certain daily activities like cooking \cite{yanai2014cooking}, restricting their applicability to diverse everyday scenarios. We aim to extend this by utilizing context-aware devices (e.g., AR smart glasses) to support open-world knowledge discovery, addressing constraints such as time and topic limitations through AI-driven technologies.

\subsection{Leveraging Machine Learning and AI to Enhance Daily Learning}
\label{sec:related_work_recent_tech_support}

Recent advances in machine learning (ML) and artificial intelligence (AI), particularly in Natural Language Processing and Large Language Models (LLMs, including Multimodal LLMs), have expanded the possibilities for informal learning~\cite{moroianu_artificial_2023}. These technologies enable systems to understand natural language queries and generate content on any topic, offering open-world knowledge and adaptability to different learning contexts \cite{yang2023dawn, brown_language_2020, radford_language_2019}. Examples of such AI-assisted informal learning systems include chatbots and visual question answering (VQA) systems \cite{Antol_2015_VQA}, which support learning both in digital contexts (e.g., simulated cultural heritage \cite{Wang_Yuan_Wang_Jiang_Zeng_2024}) and physical settings (e.g., daily objects \cite{dogan_augmented_2024, lee2024gazepointAR, wang2024G-VOILA}, everyday activities \cite{google_project_2024}, and museums \cite{Rachabatuni_Principi_Mazzanti_Bertini_2024, Bongini_Becattini_Bagdanov_Del_Bimbo_2020}). These systems are excellent for intentional learning, where users actively seek knowledge, but they often lack support for incidental learning, which we aim to address.   

To enhance user experiences, LLM-based systems increasingly incorporate personalization through individual user profiles or group personas \cite{xu2024can, sun_building_2024, ha_clochat_2024, Richardson2023Summarization}. Personalized systems have been shown to increase engagement and motivation for informal learning \cite{Draxler_Brenner_Eska_Schmidt_Chuang_2022, chen2015personalized}, yet many LLM/MLLM-based informal learning tools still lack personalization capabilities (with notable exceptions like the proposed G-VOILA framework \cite{wang2024G-VOILA}, yet without implementing this feature for real-world evaluation).
While we envision integrating personalization with LLM-based systems could enhance informal knowledge generation, how to accommodate both long-term preferences and real-time context remains underexplored, especially for incidental learning scenarios where users lack explicit learning intentions. In this work, we investigate how personalized MLLM-based systems can leverage context awareness to enhance the informal learning experience during daily activities.

\subsection{Wearable AI Assistants for Everyday Learning}
\label{sec:related_work_head_up_computing_tech_support}

Wearable devices offer greater convenience and continuous use compared to traditional computing devices across various domains~\cite{jhajharia_wearable_2014}. They enable new paradigms, such as wearable computing \cite{cliff_wearable_2005} and heads-up computing \cite{zhao2023headsup}, which provide real-time assistance during daily activities, including learning \cite{chu_research_2023, almusawi_wearable_2021, janaka_tom_2024}. In particular, AR/MR HMDs (a.k.a. AR smart glasses) display digital information overlaid on the physical world \cite{billinghurst_survey_2015, azuma_survey_1997}, enhancing context-aware learning experiences \cite{kumar_use_2018, Caetano_Lawson_Liu_Sra_2023, weerasinghe_vocabulary_2022, loh_augmented_2019, Vazquez_Nyati_Luh_Fu_Aikawa_Maes_2017}.

The integration of LLM/MLLM-based AI into wearable devices, coupled with in-situ context sensing \cite{engel2023project, grauman2022ego4d}, unlocks new opportunities for informal learning \cite{lee2024gazepointAR, google_project_2024, wang2024G-VOILA, janaka_tom_2024}. For example, Google Project Astra \cite{google_project_2024} allows users to learn about physical objects and activities through natural interactions (e.g., voice or gesture queries) using AR HMDs. Similarly, GazePointAR \cite{lee2024gazepointAR} and G-VOILA \cite{wang2024G-VOILA} use gaze-assisted voice queries to facilitate knowledge acquisition in situ. While these systems facilitate intentional learning on any topic, they do not address incidental learning. For instance, GazePointAR \cite{lee2024gazepointAR} and G-VOILA \cite{wang2024G-VOILA} utilize gaze to identify ``areas of interest/referential pointing'' but rely on \textbf{explicit verbal queries to retrieve desired information}. In contrast, adopting a ``desire prediction'' approach \cite{li2024omniactions, pandalens24cai, chang2024worldscribe}, we aim to leverage gaze data not only to detect focus levels on environmental entities but also to infer user mental states and implicit learning desires through user profile analysis. Additionally, we strive to extend beyond objects directly gazed at, delivering valuable knowledge about peripheral or overlooked entities at opportune moments. This query-free, context-aware approach facilitates seamless incidental learning.

In this work, we introduce \AiGet{}, a proactive wearable AI assistant that supports both AI-initiated and user-initiated multimodal interactions for informal knowledge acquisition in daily life. Additionally, we explore how the continued use of AI-assisted informal learning, specifically for incidental learning, impacts learning experiences across diverse scenarios.

\section{Study Overview}

Our research began with a formative study to understand users' knowledge acquisition needs in daily life. Subsequently, we developed the proof-of-concept system, \AiGet{}. We then conducted an in-lab evaluation to assess \AiGet{}'s ability to generate desirable knowledge and a real-world study to evaluate its feasibility and collect additional design insights. 
\textbf{All studies} were approved by our university's institutional review board (IRB), and participants were compensated $\approx$ 7.5 USD per hour, a standard rate for user studies in the local context.

\section{Formative Study: Understanding Informal Learning Desires during Daily Activities}

\label{sec:study1:RQ}
We envision a wearable AI \assistant{} that proactively provides relevant, in-situ knowledge based on the user's environment. However, a key question remains: \textit{\textbf{RQ:} What features do users expect from a wearable AI \assistant{} to support their knowledge acquisition during daily activities?} 

To address this question in this formative study, we explore the main barriers that prevent users from discovering informal knowledge in their everyday environments, identify the types of knowledge users desire, and conclude with four design goals to satisfy users' expectations for informal learning with wearable AI \assistant{s}.

\subsection{Participants}
We recruited twelve volunteers (P1-P12, 7 females, 5 males, Age: \meansd{24.9}{3.8} years) from the university community, all self-reporting professional working fluency in English. To ensure the accuracy of our eye-tracking equipment, we selected participants with normal or corrected vision, excluding those wearing spectacles.
Among the twelve participants, six reported using VQA apps or search engines (e.g., Google Lens) to query objects/entities in daily life at least twice per day, five used such tools with medium frequency (3-7 times per week), and one participant rarely used these technologies.

\subsection{Apparatus}
\label{sec:study1:apparatus}

Following prior research \cite{pandalens24cai, blum2006insense, buzova2020multisensory_travel}, to capture users' natural behavior during their daily routines, we utilized portable devices to record users' visual experiences and behaviors. Participants wore a backpack containing a laptop (MacBook Pro 14-inch with M2 Pro chip), which collected all the recordings. Their visual experiences, including gaze patterns from a first-person view (FPV), were recorded using a Pupil Core\footnote{\url{https://pupil-labs.com/products/core/}} eye tracker (World Camera: 30Hz, 1080p, FoV: 139\textdegree{}$\times{}$83\textdegree{}; Eye Cameras: 120Hz) connected to the laptop. An accompanying experimenter, maintaining a distance to avoid interference while ensuring participants' safety, recorded participants' actions using a mobile phone (iPhone 12) from a third-person view (TPV). This TPV feed was streamed via Zoom. Zoom synchronized and recorded the FPV, gaze, and TPV, enabling playback for reviewing recorded experiences later.

\subsection{Study Design and Procedure}
To explore user behavior and learning desires in daily contexts, we designed a study that recorded participants' routines across three settings: casual walking in indoor campus areas, outdoor campus areas, and shopping. These settings allowed us to capture informal learning opportunities in environments with both familiar (e.g., buildings on campus) and unfamiliar entities (e.g., stores with imported products). They also represented a range of information density: low (e.g., outdoor campus), medium (e.g., indoor campus with posters and exhibitions), and high (e.g., stores with a wide variety of products). 

The study was conducted in three phases:
\textbf{1) Pre-Study Interview:} Participants were interviewed to understand their daily routines, informal learning habits, challenges, and desired learning goals.
\textbf{2) Natural Behavior Recording:} Participants followed a designated path (indoor to outdoor campus to a market with unfamiliar imported products) while their behaviors were recorded using the apparatus (Sec~\ref{sec:study1:apparatus}), without experimenter intervention.
\textbf{3) Think-Aloud and Reflection:} Participants retraced the path, thinking aloud about the knowledge they wanted and why. After completing the path, they reviewed their FPV and gaze recordings at 2x speed, annotating desired knowledge points \cite{janaka_pilotar_2024}. To understand if any knowledge was ignored yet desirable, a “Wizard of Oz” simulation using GPT-4V provided overlooked knowledge for familiar and unfamiliar entities (e.g., animals, plants, products) and both attended and ignored items. Participants' expectations for wearable AI interactions were also documented.

\subsection{Data Analysis}
\label{sec:study1:analysis}

Upon consolidating 12 transcribed interview notes with observational notes detailing user behaviors and environments (i.e., desire-to-know moments with remarks, including what and why to learn, challenges with prior learning, and envisioned AI assistance), we employed a thematic analysis as outlined by Braun and Clarke \cite{braun_using_2006} (detailed in Appendix~\ref{appendix:study1_analysis}).

\subsection{Findings}
\label{sec:study1:results}

\subsubsection{Four Common Barriers Limiting Users from Discovering Informal Knowledge in Daily Activities}
\label{sec:study1:results:rq1}
Participants, primarily university students and staff, described weekday routines centered on commuting, work, and occasional sports, with weekends involving more personal activities like shopping and traveling. Within these routines, participants identified four main barriers to learning in their daily environments. 
(1) \textit{Time and Attention Limitation} \cite{bruya2010effortless, baron2023thinking}: Aligning with informal learning literature \cite{marsick_informal_2001, livingstone_adults_2001}, all participants mentioned this as the most significant barrier. They are aware of missing out on new information because they focus on existing tasks and feel that the effort required to seek out new information outweighs the potential benefits, leading to "Known Unknowns"\footnote{Things we are aware of but don't understand, from Rumsfeld Matrix \cite{saravanan_rumsfeld_2021}}.
(2) \textit{Inattentional Blindness} \cite{mack2003inattentional}: Participants often failed to notice interesting elements in their environment due to cognitive constraints, such as being preoccupied with other tasks or work-related thoughts, leading to "Unseen" knowledge.
(3) \textit{Illusion of Explanatory Depth} \cite{rozenblit2002illusion_explanatory_depth}: Participants overestimate their understanding or familiarity with surrounding entities, leading to "Unknown Unknowns"\footnote{Things we are neither aware of nor understand} unless pointed out by others (e.g., \quoteby{P2}{I never knew the commonly seen tree is the national tree here and people used it for cutting boards.}
(4) \textit{Uncertainty Avoidance} \cite{money2003uncertainty_avoid}: Barriers like language gaps lead users to avoid exploring unfamiliar environments or subjects (e.g., \quote{unfamiliar imported food in stores}), also resulting in "Known Unknowns"\footnote{Things we are aware of but don't understand}. 
Although these barriers hinder knowledge discovery in daily life, P6 noted that they do not signify a loss of curiosity. Instead, they tend to make quick assumptions about unfamiliar things (e.g., unfamiliar fruit taste) to save time, even if those assumptions are sometimes incorrect.

Meanwhile, participants have shifted their primary informal knowledge acquisition to passive consumption through social media platforms like Twitter and YouTube, aligning with prior research \cite{oeldorf2018social_media_engagment, haythornthwaite2022analytics_social_media}.
This behavior contrasts with their past, particularly in childhood, when they actively questioned their surroundings. Now, they only seek in-situ knowledge under specific conditions: when highly interested and with ample free time (e.g., \quoteby{P9}{I used plant recognition apps to search for very unique flowers I encountered}), or when the information has high practical value (e.g., \quoteby{P2}{I will search for tips before traveling or going to exhibitions}).

\subsubsection{Desired Types of Knowledge for Addressing Informal Learning Barriers}
\label{sec:study1:results:rq2}

To overcome \textit{Time and Attention Limitation}, participants expressed a desire for proactive AI to provide knowledge with little effort during daily moments. They emphasized that such a system should offer contextually relevant knowledge. Revised Bloom's Taxonomy \cite{krathwohl2002revision} categorizes four types of knowledge: Factual, Conceptual, Procedural, and Metacognitive (see details in Appendix~\ref{appendix:knowledge_examples}). During daily activities, Metacognitive knowledge is less desirable, as it demands complex thinking and could exacerbate \textit{Time and Attention Constraints}. As P8 noted, \quote{If it prompts me to think about milk safety in different containers while shopping, it's too much. I'd rather review that information later.}

For the remaining three categories, we found that their preferences are linked to the entity's familiarity level, echoing the need to address \textit{Illusion of Explanatory Depth} and \textit{Uncertainty Avoidance} (Sec~\ref{sec:study1:results:rq1}). 

For \textit{Commonly Seen or Familiar} entities that lead to \textit{Illusion of Explanatory Depth}, participants preferred \textbf{interesting factual knowledge offering surprises} (e.g., \quoteby{P10}{shocking facts or statistics about environmental issues associated with plastic Coca-Cola bottles}) or \textbf{useful procedural tips} relevant to ongoing or future tasks (e.g., \quoteby{P7}{how to mix common beverages for unique flavors}). 
For \textit{Unfamiliar} entities, which lead to \textit{Uncertainty Avoidance}, participants found \textbf{basic introductions (factual knowledge)} in an easy-to-understand format most helpful. Additionally, P3 and P4 appreciated when unfamiliar entities were \textbf{linked to familiar concepts}, making new information more relatable and easier to grasp.

We also found that knowledge preference regarding the \textit{Focus Level of the Entity}, i.e., whether the provided knowledge relates to what users are actively observing, can partially mitigate the effects of \textit{Inattentional Blindness}. These preferences were related to observed three gaze and motion patterns, similar to prior literature \cite{janaka_visual_2022}, revealing users' varying interests based on their environmental engagement and focus:

(1) When users show a \saccade{} gaze pattern (i.e., random gaze movements across different entities in FPV) during casual walking, their mental state often suggests that they are looking for something interesting, relaxing their mind, or thinking about unrelated tasks. In these cases, participants appreciated fun, factual knowledge about their surroundings, \textbf{regardless of whether the associated entity is scanned} (i.e., \quoteby{P4}{unnoticed interest is also welcomed}). However, four participants emphasized that if they were deep in thought, which is not reflected in gaze patterns alone, they would likely turn off the \assistant{} to avoid interruptions.

(2) During \quickBrowse{} (i.e., rapid gaze scanning of related entities in FPV), where users scan familiar entities (e.g., store shelves), they have a medium interest level in the scanned target and typically are seeking something more engaging. In this context, providing interesting \textbf{knowledge related to the scanned entities} can enhance users' interest in familiar items.

(3) When users are fully \focused{} (i.e., gaze fixations on specific entities in FPV), it indicates high interest, often because the entity is unfamiliar, they want to learn more about it, or they are making a decision. In such moments, participants found it helpful to receive factual and conceptual knowledge that \textbf{builds connections or comparisons between the primary focused entity and similar nearby entities}. 
They also appreciated comparisons between the entity and something familiar, even if not physically present. For example, when P3 found pre-made breakfast in the store, they wanted to know what made these options unique compared to similar ones in their hometown. Such comparisons provide a familiar context, making it easier for users to understand and engage with new information.

\subsubsection{Summarized Design Requirements for Wearable AI Assistant to Help Users Learn from Environment} 
\label{sec:study1:results:rq3}
\label{sec:study1:design_goals}

Given the identified barriers, user preferences for knowledge acquisition, and considering the unique challenges of information intake in mobile settings, we synthesized four design requirements for a wearable AI \assistant{} based on participant feedback.

\paragraph{\designGoalContextUnderstanding{}: Analyze Context and Provide ``Unseen'' and ``Unknowns'' in Real Time}
Building on the insights for varying knowledge desires under different contexts, participants envision the AI \assistant{} analyzing their environment and gaze behavior to predict their primary activities and potential learning desires, subsequently providing knowledge support about their ``unnoticed'' or ``unknown interesting knowledge'' about the in-situ environment. In addition to physical contexts, the \assistant{} could check digital contexts, like the ``agenda'', to predict support for follow-up activities, such as offering guidance on suitable food and drinks before exercising when shopping. In addition, the participants mentioned the need to support human initiative queries in case \quoteby{P1}{the AI doesn't know what exactly I want to know.}

\paragraph{\designGoalPersonalization{}: Personalization is Needed to Fine-Tune Knowledge Content}
Participants emphasized the need for personalization, noting that it enables the AI \assistant{} to understand their long-term interests and familiarity level with various entities in their environment. By assessing familiarity, the \assistant{} can tailor knowledge types and expertise levels (Sec~\ref{sec:study1:results:rq2}). To achieve this, they suggested collecting a \quoteby{P3}{basic user profile like social media platforms did when registering}. While personalization is valued, participants also expressed concerns about \quoteby{P2}{the system constantly suggesting information based on past interests, similar to social media algorithms}, which could become annoying. Instead, they hope the \assistant{} utilizes this familiarity to introduce new perspectives and noteworthy entities in daily activities that might otherwise be overlooked.

\paragraph{\designGoalPrioritization{}: Knowledge Prioritization to Provide the Most Valuable Information}
\label{sec:study1:results:rq3:prioritization}
Participants appreciated discovering information from their environment but emphasized the need to prioritize the most valuable content, avoiding information that is trivial. To address this, we propose a multi-factor weighting system to prioritize knowledge selection. First, \textit{Novelty} was considered crucial, with most participants (10 out of 12) agreeing that the knowledge \textbf{must} be new to them, either by introducing entirely new perspectives or by adding unknown details to common knowledge. For example, instead of simply stating that "palm trees provide habitats for various species," the \assistant{} could highlight that "palm trees provide habitats for species like the Palm Weevil, which lays its eggs in the trunk." Additionally, participants identified three factors as beneficial, though not always necessary: (1) \textit{Alignment with Personal Interests and Values}, such as providing bargain-hunting tips for participants who value ``saving''; (2) \textit{Usefulness} for ongoing or future tasks, like helping health-oriented users compare sugar levels when buying drinks; and (3) \textit{Unexpected Perspectives} that offer surprising insights or correct misconceptions, such as "zero-sugar drinks do not mean they contain zero calories."

These priorities align with existing literature on user information-seeking behavior, which emphasizes information utility \cite{kelly_individual_2021, sharot2020utility_decision_making}, and are consistent with measures of serendipity in recommender systems \cite{kotkov2023serendipity}, focusing on novelty, unexpectedness, and relevance. By integrating these elements into the weighting formula: \PrioritizationFormula{}, the \assistant{} can prioritize the most valuable knowledge for users.

\paragraph{\designGoalMinimizedDistraction{}: Minimize Distraction to Primary Tasks in Mobile Settings}
Given the use of the wearable \assistant{} in scenarios where users have other primary tasks and consume information on the move, participants highlighted two main points to minimize distraction. First, regarding knowledge delivery, participants emphasized the need to minimize the cognitive load during receptive moments for learning. As P7 noted, \quote{the content should be concise and easy to digest, like a podcast.} They also emphasized that the \assistant{} should provide only 1-2 pieces of information at a time to prevent overwhelming users, especially during tasks requiring focus, unless they explicitly requested more. Second, concerning system control, participants emphasized the importance of accommodating moments when users are unwilling to receive knowledge. This includes allowing users to turn the \assistant{} off as needed for extended periods (e.g., while rushing to a destination or preoccupied with work or study) or easily interrupt single information delivery if the content is unsatisfactory. These considerations aim to ensure that the \assistant{} enhances the user experience rather than hinders the user's primary activities in mobile settings.

\begin{figure*}[hptb]
\centering
\includegraphics[width=1\linewidth]{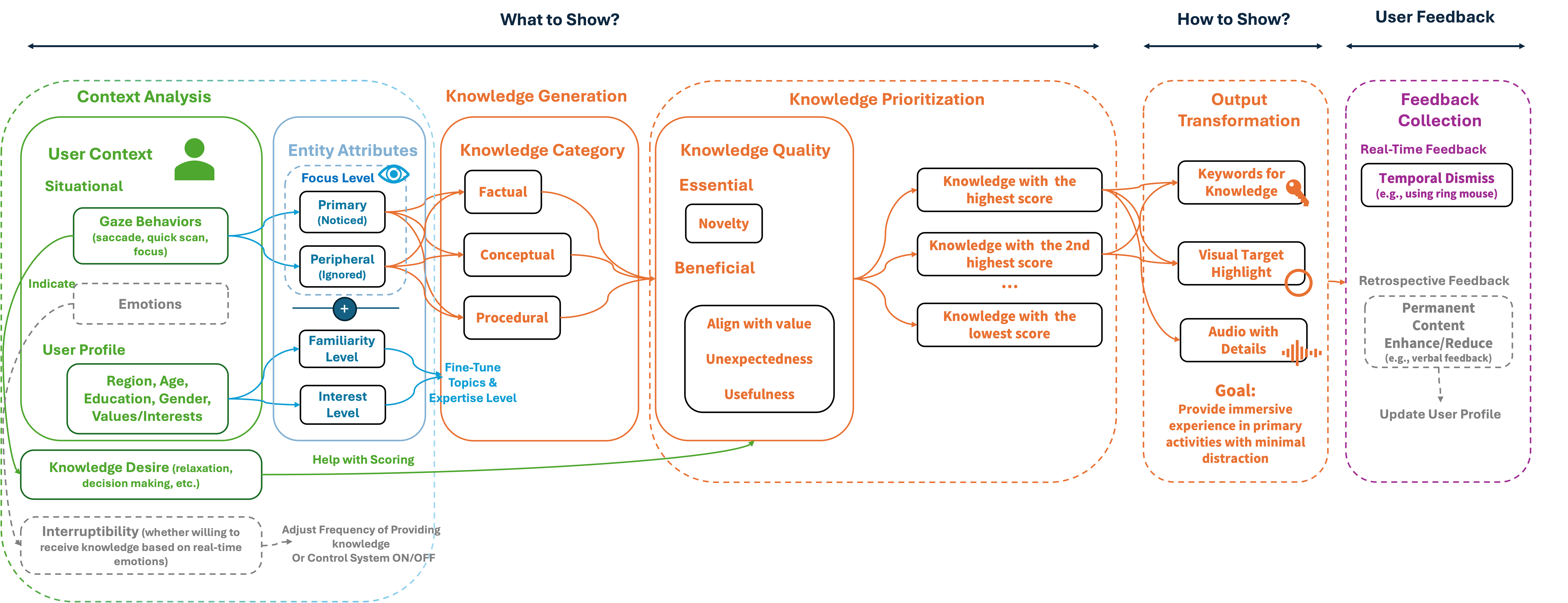}
\caption{Proposed Framework for Generating Desired Knowledge from Daily Moments. Note: Gray dashed boxes and lines represent user suggestions for future system improvements, collected from the final real-world study (Sec~\ref{sec:discussion:future_improvements}) and not implemented or evaluated in the current paper. Our implementation, \AiGet{}, is detailed in system design (Sec~\ref{sec:system}).}
\Description{The diagram presents a framework for generating knowledge from daily moments. It starts with Context Analysis, which considers gaze behaviors, emotions, and user profiles (e.g., age, education, values) to determine focus, familiarity, interest, and interruptibility based on the user’s emotional state. Entity Attributes are assessed to score the relevance of information based on whether the object is in primary or peripheral focus, and the user's familiarity and interest levels. Knowledge Generation then categorizes knowledge into factual, conceptual, or procedural, fine-tuning it to the user’s expertise. In Knowledge Prioritization, information is ranked based on qualities like novelty, unexpectedness, usefulness, and alignment with user values. Output Transformation delivers the knowledge through keywords, visual highlights, and audio feedback, minimizing distractions. Lastly, Feedback Collection allows users to dismiss content or provide feedback to update their profile. Gray dashed boxes indicate user suggestions from real-world studies, such as adjusting delivery frequency, which are not yet implemented.}
\label{fig:study1:framework}
\end{figure*}

\vspace{3mm}
Based on these design requirements, we propose a framework for desired knowledge generation, as illustrated in Figure~\ref{fig:study1:framework}. This framework outlines the process of transforming daily moments into valuable learning opportunities while minimizing distractions and incorporating user feedback. 

\section{\AiGet{} System}
\label{sec:system}

We introduce \AiGet{}, a \textbf{proof-of-concept} system that aligns with the aforementioned design goals (Sec~\ref{sec:study1:design_goals}). In this section, we detail the primary features of the \AiGet{} system, its implementation, and an in-lab simulated ablation study conducted to evaluate its knowledge generation capability.

\subsection{Key Features of \AiGet{} System}
\label{sec:system:key_features}
\begin{figure*}[hptb]
\centering
\includegraphics[width=1\linewidth]{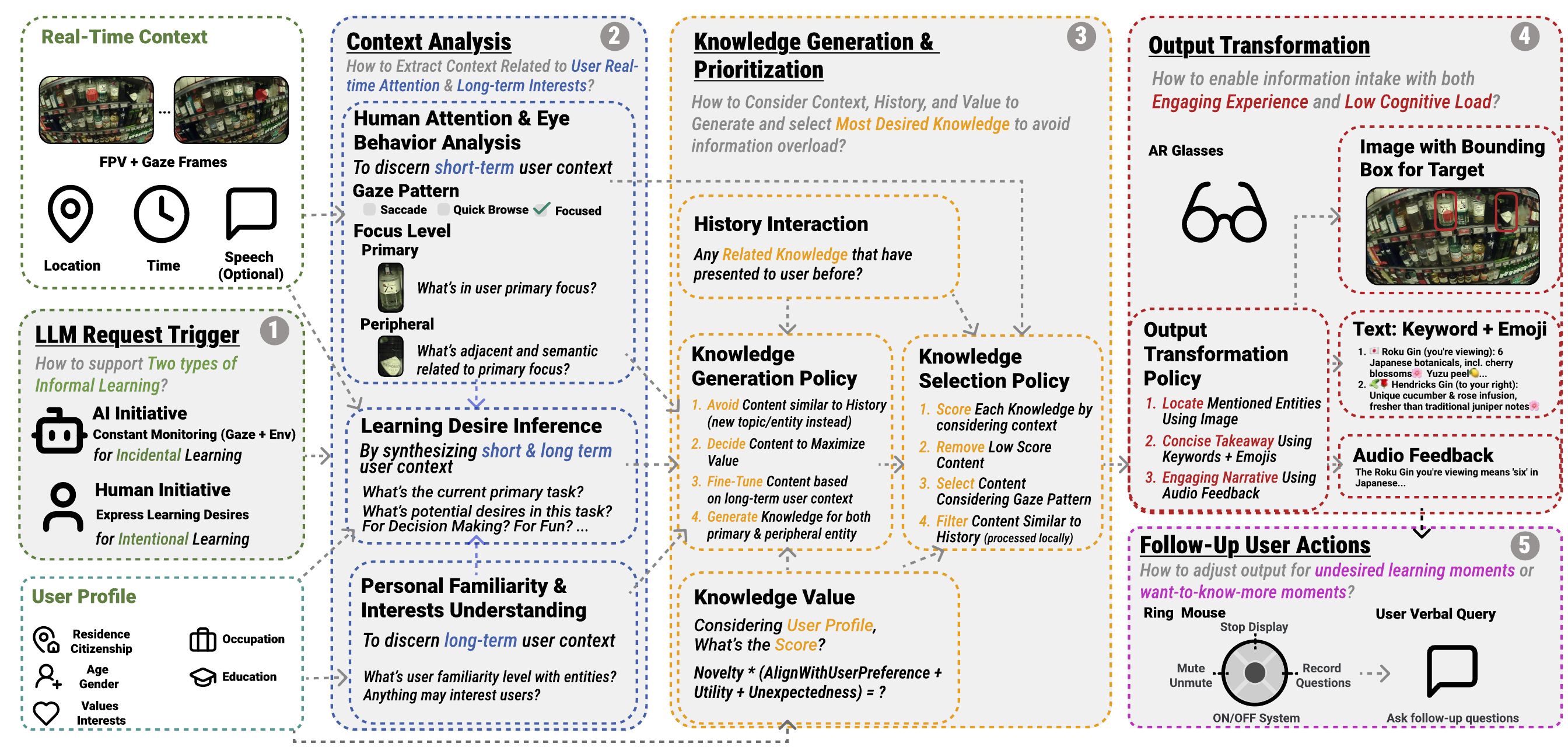}
\caption{\added{\AiGet{}'s System Processing Pipeline. (1) LLM Request Trigger: Supports mixed-initiative knowledge queries. (2) Context Analysis: Analyzes real-time user attention and long-term interests and infers learning desires. (3) Knowledge Generation \& Prioritization: Filters redundant content, prioritizing the most valuable knowledge to avoid overload. (4) Output Transformation: Presents knowledge in a multimodal format to balance engagement and cognitive load. (5) Follow-Up User Actions: Enables user control over interaction. Some details are omitted in the figure. Full details and agent prompts are in Appendix~\ref{appendix:prompt_for_llm}.}}

\Description{It illustrates the system processing pipeline of AiGet, highlighting the four key modules and their interactions: 1. Real-Time Context: Displays inputs such as FPV and gaze frames, location, time, and optional speech. 2. LLM Request Trigger: Supports two modes: (a) AI Initiative for incidental learning through gaze and environmental monitoring and (b) Human Initiative for intentional learning based on expressed desires. 3. Context Analysis: Breaks down user context into short-term (e.g., gaze patterns and primary/peripheral focus) and long-term (e.g., user profile values and interests). 4. Knowledge Generation and Prioritization: Combines historical interaction, content generation policies, and selection policies to optimize knowledge output. It emphasizes novelty, alignment with user preferences, and minimizing redundant or low-value information. 5. Output Transformation: Adapts knowledge delivery to AR glasses using multimodal outputs, including text with emojis, images with bounding boxes, and audio feedback, all designed for low cognitive load and engaging interaction. 6. Follow-Up User Actions: Details user controls, such as using a ring mouse to mute/unmute, stop displays, or initiate follow-up queries.}
\label{fig:study2:sys_flow}
\end{figure*}

The \AiGet{} system pipeline, as shown in \textbf{Figure~\ref{fig:study2:sys_flow}}, consists of five key stages: (1) LLM Request Trigger, (2) Context Analysis, (3) Knowledge Generation and Prioritization, (4) Output Transformation, and (5) Follow-up User Actions. It was developed following the design goals (Sec~\ref{sec:study1:design_goals}) and (iterative) pilot testing with eight users from the university, all with backgrounds in UI/UX design or HCI.

\subsubsection{LLM Request Trigger (\designGoalContextUnderstanding{})}
\label{sec:system:llm_request_trigger}
Aligning with \designGoalContextUnderstanding{}, to support both proactive AI prompting and user-initiated queries to discover ``unseen'' and ``unknowns'', \AiGet{} adopts a mixed-initiative approach~\cite{horvitz1999principles, amershi_guidelines_2019} to trigger LLM requests (Figure~\ref{fig:study2:sys_flow}~(1)).

\paragraph{AI-Initiative: Constant Sensing} 
Since users may not always notice all the interesting entities in their environment and be aware of the ``unknown knowledge''  associated with them, \AiGet{} continuously monitors the environment (via FPV camera) to identify hidden yet desired knowledge from surroundings. To prevent information overload and repetition, LLM requests are triggered only when both a minimum time interval (12 seconds) has passed and a significant FPV difference threshold is met (see Appendix~\ref{appendix:system:fpv_diff} for technical details).  This ensures users have enough time to process each piece of generated knowledge on mobile situations.

\paragraph{AI-Initiative: Implicit Gaze Fixation} 
In addition to constant sensing, \AiGet{} monitors the user's gaze pattern to trigger LLM requests. When \AiGet{} detects user fixation patterns (eyes focused on a small area, deviating no more than 4.91 degrees for at least 1 second, as suggested by prior research \cite{pandalens24cai}), it indicates potential interest in the target object and triggers an LLM request.

\paragraph{User-Initiative: Explicit Verbal Query} 
Recognizing that users may have specific questions during daily activities or may want to ask follow-up questions after being inspired by AI's proactive suggestions (i.e., intentional learning), \AiGet{} supports explicit verbal queries from users, similar to the prior visual Q\&A system \cite{wang2024G-VOILA, lee2024gazepointAR}. To prevent false triggers in noisy environments, users can press the right button on the handheld ring mouse to control their query~\cite{pandalens24cai}. The system then incorporates the user's verbal comments or questions into the LLM request.

\subsubsection{Context Analysis (\designGoalContextUnderstanding{}, \designGoalPersonalization{})}
\label{sec:system:context_analysis}
When an LLM request is triggered, \AiGet{} compiles multimodal contextual data, including the user's FPV image frames with gaze data, time, location, and any user query transcriptions for analysis.

To support \designGoalContextUnderstanding{}, \AiGet{} employs a \textbf{Context Analysis Agent} (Figure~\ref{fig:study2:sys_flow}~(2)) to extract meaningful contexts by analyzing short-term attention and long-term interests. This enables the system to predict learning desires and minimize the generation of irrelevant or low-quality knowledge from extensive raw multimodal data, particularly when explicit user learning desires are not expressed.
Specifically, the agent recognizes user activity, labels entities, predicts user intentions, and identifies potential unseen or unknown entities and knowledge topics aligned with the user's interests.

To help the LLM understand the user's primary activities and gaze patterns, \AiGet{} overlays gaze positions on each FPV frame with red circles, enabling LLM to discern focused regions while maintaining global context \cite{Shtedritski_2023_ICCV, yang2023dawn}. The LLM then analyzes the user's activities, identifies objects in both primary and peripheral visual fields, and interprets gaze patterns (Sec~\ref{sec:study1:results:rq2}) using predefined rules: \saccade{} (random gaze movements across different entities in different frames), \quickBrowse{} (rapid scanning of related objects in different frames), and \focused{} (sustained attention on specific objects in most frames). The LLM also predicts potential learning desires based on these patterns and contextual information, such as ``looking for a specific keyboard model or comparing different keyboard options'' when users focus on various keyboards in stores.

To enhance analysis and support \designGoalPersonalization{}, the system provides the LLM with a \textbf{user profile} \added{(Figure~\ref{fig:study2:sys_flow} (User Profile))} \cite{xu2024can, wang2024G-VOILA} containing values, interests\footnote{To ensure broad coverage of user interests and values, we followed a two-step process: (1) Curating 14 informal knowledge topics (e.g., Animals \& Plants) and 58 value categories (e.g., Physical Value-Beauty) from taxonomies in prior literature \cite{livingstone2001adults, marino2019ok, SCHWARTZ19921, personal_values} to guide participant selection; (2) Encouraging participants to expand the list with detailed topics (e.g., "Plant Care Tips") that aligned with their experiences and learning desires.}, and basic demographic information \added{(i.e., age, gender, nationality, residence, and academic/professional background; see Appendix~\ref{appendix:system:user_profile} for a sample profile)}. \deleted{A sample profile is as follows: \quote{{"Values/Interest": ["healthy life", "fitness", "flowers", "coffee lover", "cat", "fun facts/history", "design"], "Age": "30", "Gender": "female", "Citizenship": "XX", "Residence": "XX", "Education": "PhD in visual design", "Occupation": "senior student in XX"}}. }Using this profile, the LLM assesses the user's familiarity with the current location and surrounding entities (e.g., "a senior student is familiar with the campus," or "coffee lovers know the basics of different types of coffee"). It also refines learning desires and identifies potential connections between the environment and the user's interests (e.g., linking the topic of "fruit nutrition" to a user's interest in a "healthy lifestyle" when passing the supermarket fruit aisle).

\paragraph{Prompt Improvement Through Iterative Design}
We identified two key challenges in accurately identifying entities from long and raw multimodal inputs in real-world scenarios: ambiguity when multiple objects overlap in gaze direction and visual recognition errors for underrepresented entities in image training datasets. To address these issues, we employed a multimodal chain-of-thought reasoning approach \cite{zhang2023multimodal}, as detailed in Table~\ref{tab:entity_identification}, which significantly improved contextual accuracy in incidental learning during our pilot tests.

\begin{table}[hptb]
\centering
\caption{MLLM Prompt Refinements to Address Entity Identification from Multimodal Inputs}
\Description{This table summarizes challenges in multimodal entity identification and solutions using MLLM prompt refinements. For Ambiguity in Gaze Interpretation, Context-Guided Selections prioritize rare elements to enhance relevance (e.g., focusing on a museum specimen rather than its jar). For Visual Recognition Errors, Multimodal-Facilitated Visual Processing uses text cues to identify underrepresented entities, such as niche products. These methods improved contextual accuracy and robustness in pilot tests.}
\label{tab:entity_identification}

\begin{tabular}{p{0.35\linewidth}|p{0.60\linewidth}}
\textbf{Challenges} & \textbf{Solutions  \& Steps} \\ \hline
\textbf{Ambiguity in Gaze Interpretation:} When gaze direction overlaps multiple entities.
\newline 
\textit{Example: Distinguishing between a glass jar and the specimen inside.} & 
\textbf{Context-Guided Selections:} 
The MLLM was instructed to describe user activity first and prioritize elements based on rarity and uniqueness to the scene and context. 
\newline 
\textit{Example: As emphasized by P1: "I can learn about tempered glass elsewhere, but the specimen is unique to the museum."} \\ \hline
\textbf{Visual Recognition Errors:} For underrepresented entities in image training dataset.
\newline 
\textit{Example: A newly released Asian drink.} & 
\textbf{Multimodal-Facilitated Visual Processing:} 
When images alone were insufficient, the MLLM performed OCR to extract textual cues in the scene, enhancing entity recognition via text labels. 
\newline 
\textit{Example: Extracting labels on bottles and shelves helped the MLLM identify the drink accurately.} \\ %
\end{tabular}
\end{table}

\subsubsection{Knowledge Generation and Prioritization (\designGoalContextUnderstanding{}, \designGoalPersonalization{}, \designGoalPrioritization{})}
\label{sec:system:knowledge_generation}

After context analysis, the \textbf{Knowledge Generation Agent} (Figure~\ref{fig:study2:sys_flow}~(3)) generates and selects the most ``desired'' knowledge candidates based on the \textit{contextual description}\footnote{It combines an FPV image with textual data from the \textbf{Context Analysis Agent} to balance visual detail retention and input token efficiency.}, interaction history, and knowledge value.

Aligning with \designGoalContextUnderstanding{} and \designGoalPersonalization{}, the agent produces factual, conceptual, and procedural knowledge on both primary and peripheral entities. The knowledge content is required to be novel, useful, and surprising, avoiding repetitive topics by checking history. The agent also fine-tunes content based on user familiarity and interest levels, e.g., offering advanced insights on familiar topics and linking new information to known concepts for unfamiliar ones.

To reduce trivial knowledge provision (\designGoalPrioritization{}), we use Chain of Thought methods \cite{wei2022chain}, instructing the LLM to score each knowledge item using the prioritization formula  (sec~\ref{sec:study1:results:rq3:prioritization}): \PrioritizationFormula{}, with each factor marked as 0 or 1 by the agent with reasoning. Only knowledge content scoring >= 2 is retained, based on pilot results. The LLM then selects the suitable knowledge based on the user's gaze pattern mode (Sec~\ref{sec:study1:results:rq2}), following rules from the formative study (e.g., for \focused{} mode, providing comparative knowledge about primary and related peripheral entities). 
No more than two concise knowledge items are presented at a time to minimize information overload.

\paragraph{Prompt Improvement Through Iterative Design} 
Designing incidental learning for mobile settings posed unique challenges in generating novel and unexpected content due to ambiguous instructions and LLM limitations in processing long contexts \cite{maharana2024evaluating}. To overcome these challenges, we refined multiple prompt strategies through iterative pilot testing, as summarized in Table~\ref{tab:novelty_unexpectedness}.

\begin{table*}[hbtp]
\centering
\caption{LLM Prompt Refinements to Address Novelty and Unexpectedness in Content Generation}
\Description{This table outlines prompt refinements for LLMs to enhance novelty and unexpectedness in content generation. To improve novelty, challenges like generic outputs and repetition were addressed using few-shot prompting tailored to user expertise and a filtering mechanism inspired by retrieval-augmented generation, ensuring the generation of fresh, relevant content. For unexpectedness, prompts incorporated a "surprise taxonomy," guiding models to focus on lesser-known facts, misconceptions, and unseen but intriguing perspectives, enhancing engagement and user interest.}
\resizebox{\linewidth}{!}{

\begin{tabular}{l|p{0.35\linewidth}|p{0.55\linewidth}}
\textbf{Aspect}          & \textbf{Challenges}                                                                                 & \textbf{Solutions \& Steps}                                                                                                                                                                                                                           \\ \hline
\textbf{Novelty} & 
\textbf{Balancing Detail with Conciseness}:
\begin{itemize}[leftmargin=*]
    \item Requirement for Conciseness led to generic outputs from naive LLMs, lacking depth for educated users.
    \item \textit{Example: "Palm trees play an important role in the ecosystem by providing food for various species."}
\end{itemize} &
    \textbf{Few Shot Prompting}:
\begin{itemize}[leftmargin=*]
    \item Employed few-shot prompting \cite{Reynolds2021fewshot} with concrete examples to guide LLMs in tailoring details to users' education levels.
    \item \textit{Example: "Palm trees emit volatile organic compounds attracting beneficial insects, reducing pesticide use."} balances conciseness with novel details, and domain experts could receive more technical precision.
\end{itemize} \\ %
& \textbf{Reducing Repetition when Revisiting Entities}:
\begin{itemize}[leftmargin=*]
    \item Content repetition undermined novelty, especially when revisiting recurring entities like trees along a route.
    \item Simply feeding full histories and requiring new content failed due to LLMs' limitations in long context processing.
\end{itemize} &
\textbf{Filtering Mechanism with Chain of Thoughts}:
\begin{itemize}[leftmargin=*]
    \item Inspired by retrieval-augmented generation (RAG) \cite{zamani_retrieval_2022}, only fed the 10 most relevant historical items based on semantic similarity* to \textit{contextual description}.
    \item Applied Chain-of-Thought method \cite{wei2022chain} to break complex tasks into smaller steps and maximize the likelihood of generating novel content:
    \begin{itemize}[leftmargin=*]
        \item Identify entities/topics already mentioned in concise history.
        \item Determine unexplored gaps.
        \item Generate content focusing on new perspectives or entities.
    \end{itemize}
    \item Post-generation, filter content highly similar to history ($\geq 0.75$)* to ensure novelty. 
\end{itemize} \\ %
\textbf{Unexpectedness}  & \textbf{Lack of Surprising Perspectives}:
\begin{itemize}[leftmargin=*]
    \item Hard to generate unexpected content without explicit guidance.
\end{itemize} &
\textbf{Specifying Potential Unexpectedness Category}:
\begin{itemize}[leftmargin=*]
    \item Guided by "surprise taxonomy" \cite{modirshanechi2022taxonomy}, prompts suggested:
    \begin{itemize}[leftmargin=*]
        \item Unseen but interesting entities (Observation-Mismatch Surprise \cite{modirshanechi2022taxonomy}).
        \item Lesser-Known Fun facts (e.g., trivia) or misconceptions (Belief-Mismatch Surprise \cite{modirshanechi2022taxonomy}).
    \end{itemize}
\end{itemize} \\ %
\multicolumn{3}{l}{\footnotesize * The similarity is calculated using the all-MiniLM-L6-v2 model, \url{https://huggingface.co/sentence-transformers/all-MiniLM-L6-v2}} \\
\end{tabular}}
\label{tab:novelty_unexpectedness}
\end{table*}

\subsubsection{Output Transformation (\designGoalMinimizedDistraction{})}
\label{sec:system:output_transformation}

To align with \designGoalMinimizedDistraction{} and pilot testing results, \AiGet{} delivers output in a multimodal format (e.g., Figure~\ref{fig:study2:sys_flow}~(4), Figure~\ref{fig:study3:berlin_wall}), balancing minimal distraction with increased engagement by reducing the cognitive load associated with information intake. The system integrates three key features: 1) Audio modality to provide full knowledge content, enhancing engagement while minimizing text reading efforts on the move \cite{vadas_reading_2006a, vadas_reading_2006b}; 2) Emojis/pictograms and text keywords to succinctly convey the content gist and enhance comprehension \cite{janaka2023icons}; and 3) Images with highlighted bounding boxes around mentioned entities to provide visual references, particularly for overlooked or unfamiliar entities. All visual information is presented in the user's peripheral vision areas, following attention-maintaining interface guidelines ~\cite{janaka_paracentral_2022, cai2023paraglassmenu}.

To optimize processing time, two agents work in parallel during output transformation, as detailed in Appendix~\ref{appendix:system:output:parallel_processing}.

\subsubsection{Follow-Up User Actions (\designGoalMinimizedDistraction{})}
\label{sec:system:followup_actions}
After receiving system output, users can perform follow-up actions using the ring mouse (Figure~\ref{fig:study2:sys_flow}~(5)). To align with \designGoalMinimizedDistraction{} and accommodate moments when users prefer not to engage, pressing the up button cancels the latest knowledge display, stopping audio and hiding visual information (\textit{Emojis and Text} and \textit{Image Reference}) when encountering unwanted information. The left button mutes or unmutes \textit{Audio feedback}, while the bottom button toggles the system on or off for extended periods, controlling proactive AI suggestions. As mentioned earlier, the right button enables users to initiate follow-up queries. The buttons and UI elements are spatially mapped for easy interaction \cite{norman_design_2013, cai2023paraglassmenu}.

\subsection{Apparatus and Implementations}
\label{sec:system:implementation}

An XReal Air\footnote{\url{https://www.xreal.com/air}} was used as the near-eye AR display, with the Pupil Core add-on for gaze detection and FPV streaming \cite{pandalens24cai}. The smart glasses were connected to a laptop (MacBook Pro 14-inch, M2 Pro chip), which served as the computing unit to display the \AiGet{} UI. A Sanwa Supply Bluetooth Ring Mouse (MA-BTRING3BK\footnote{\url{https://www.sanwa.co.jp/product/syohin?code=MA-BTRING3BK}}) was used to control the UI.

Python was used for both the frontend and backend, utilizing a TKinter-based interface to seamlessly handle real-time capture and concurrent processing of various context data and user interactions. For the LLM pipeline, we employed the Gemini-series models (i.e., Gemini-1.5-Flash for contextual analysis and output transformation, and Gemini-1.5-Pro for knowledge generation). These models were selected due to the support of 1) large context windows, enabling the system to process multiple contextual inputs and maintain a long knowledge history, and 2) safety settings to avoid potential ethics issues with generated content. 
Detailed system implementation and prompt information can be found in Appendix~\ref{appendix:system:implementation} and Appendix~\ref{appendix:prompt_for_llm}.

\subsection{In-Lab Evaluation of \AiGet{}'s Knowledge Generation Pipeline}
\label{sec:in_lab_study}
To verify whether the \AiGet{} pipeline can generate desired knowledge using multimodal contextual data and to determine which aspects of our design goals (Sec~\ref{sec:study1:design_goals}) contribute to such knowledge generation, we conducted an in-lab simulated evaluation.

Specifically, we conducted an ablation study by removing \textbf{two key components} of \AiGet{} pipeline---knowledge generation rules (\designGoalPrioritization{}) and personalization (\designGoalPersonalization{})---from the pipeline, following the method used in prior research \cite{xu2024can, wang2024G-VOILA}. The pipeline originally consisted of a Multimodal LLM, User Profile, and Rules (Prioritization with Focus/Familiarity Analysis). We kept the Multimodal LLM (\designGoalContextUnderstanding{}) as it is compulsory for context-relevant knowledge generation. Then, we compared the knowledge generated by the three resulting pipelines to understand what components and associated guidelines affect the desirability of knowledge. Note: \designGoalMinimizedDistraction{} is not considered here, as it focuses on interaction rather than knowledge generation.

\subsubsection{Two Baselines}

\paragraph{\BaselinewoR{} (i.e., Multimodal LLM + Profile)} 
This baseline omits the \textbf{R}ules that associate knowledge categories with gaze pattern and familiarity analysis while retaining the user profile in the prompt. The LLM is instructed to analyze multimodal input data (FPV + gaze overlay, location, and time) and generate knowledge that `enhances interest, expands knowledge, and includes serendipitous information' following our original objectives (Sec~\ref{sec:system}), tailored to the user's profile when possible. It's directed to produce no more than two knowledge items, avoiding basic information that neither broadens the user's perspective nor provides utility (but is not required to calculate specific scores as \AiGet{} did). To assess whether basic user profiles could lead to more personalized and valuable knowledge, both \AiGet{} and \BaselinewoR{} use identical user profiles, as mentioned in Sec~\ref{sec:system:context_analysis}.

\paragraph{\BaselinewoRP{} (i.e., Multimodal LLM only)} This version removes both the \textbf{R}ules and the \textbf{P}ersonalized user profile. Similar to \BaselinewoR{}, but without the benefit of a user profile, this pipeline generates knowledge based solely on multimodal input data. It follows the same guidelines for selecting relevant knowledge and avoiding basic information.

To ensure a fair comparison, all baselines maintain a similar prompt structure, prompt engineering methods, and LLM models for knowledge generation as \AiGet{}, except for the aforementioned differences. Note that we didn't implement a version that ``only removes personalization while keeping the rules'', as the rules depend on user profiles. Thus, both components are omitted in \BaselinewoRP{}, aligning with the prior research's method~\cite{xu2024can}. Detailed prompts for both baselines are in Appendix~\ref{appendix:prompt:baselinewor} and \ref{appendix:prompt:baselineworp}.

\subsubsection{Participants} 
As the perception of generated knowledge is subjective and context-dependent, to have a realistic comparison between the \AiGet{} pipeline and baseline pipelines, participants rated the knowledge generated by three pipelines using their own recorded daily activities, where they had first-hand experience. 

Thus, we invited 12 participants from the formative study to evaluate knowledge generated from their previous daily activity recordings. This approach enabled personalized evaluation based on participants' actual experiences, allowed them to judge knowledge accurately, and improved resource efficiency by eliminating the need for new recordings.

\subsubsection{Study Design and Data Preparation}

A within-subject design was used to compare the knowledge generation quality across different pipelines. Each participant rated knowledge generated from 9 moments, with 3 knowledge items per moment (one from each pipeline), totaling 27 knowledge items per participant. 

\textit{Moments Selection.} 
For each scenario from the formative study (i.e., indoor and outdoor casual walking on campus and shopping), three moments with different scenes (selected based on FPV similarity, detailed in Sec~\ref{sec:system:llm_request_trigger} and Appendix~\ref{appendix:system:fpv_diff}) were randomly chosen to balance variety and avoid overwhelming participants when evaluating multiple aspects of each pipeline. Although not all possible generated knowledge items were assessed, the selected samples realistically estimated each pipeline's performance. In total, 108 data samples were selected for evaluation (3 moments per scenario $\times$ 3 scenarios per participant $\times$ 12 participants).

\begin{figure*}[hptb]
\centering
\includegraphics[width=1\linewidth]{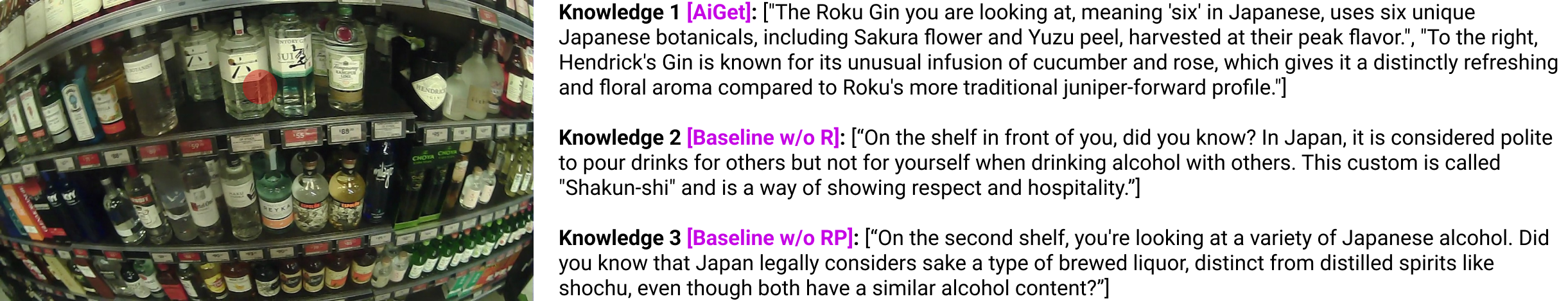}
\caption{An example of comparative knowledge from three MLLM pipelines. Note: Participants can't see each pipeline's name (in purple).}
\Description{This image shows three rows of comparative knowledge output from different MLLM (Multimodal Large Language Model) pipelines with an FPV frame of alcohol shelf on the left side. Each row contains a set of knowledge statements. The first pipeline highlights product information, including ingredients and unique characteristics of Japanese gin. The second pipeline shares cultural etiquette for pouring drinks in Japan, while the third describes Japanese alcohol categorization, distinguishing between sake and shochu. The pipeline names (in purple) are hidden from participants during evaluation.}
\label{fig:study2:tech_eval_doc}
\end{figure*}

\textit{Knowledge Generation.} 
For each moment, knowledge was generated using all three pipelines and presented in randomized order with anonymous ID (i.e., Knowledge 1-3) to ensure unbiased evaluations. See Figure~\ref{fig:study2:tech_eval_doc} for an example.

\subsubsection{Measures} 
\label{sec:study2:measures}
We assessed the desirability of the generated knowledge across several key aspects using subjective measures, following prior research \cite{kotkov2023serendipity, kelly_individual_2021, sharot2020utility_decision_making, Sommerauer_Müller_2014}, including \novelty{}, \personalization{}, serendipity (encompassing \usefulness{}, \unexpectedness{}, and \relevance{}), \interesting{}, and \deepenKnowledge{}---each measured using a 7-point Likert scale (1 for ``Strongly Disagree'' and 7 for ``Strongly Agree''). For details, see Appendix~\ref{appendix:measures}.

Additionally, to evaluate whether the information was appropriate for presentation during primary tasks \cite{pandalens24cai, cai2023paraglassmenu, janaka_glassmessaging_2023}, we collected \overallScore{} for receiving such knowledge during primary tasks, \notAnnoying{}, and \notOverload{}, also using 7-point Likert scales.

Moreover, we also recorded instances where the MLLM pipeline did not generate any knowledge because it determined nothing valuable to provide (Note: for these moments, participants only rated three measures: \overallScore{}, \notAnnoying{}, and \notOverload{}).

\paragraph{Analysis}
A Wilcoxon signed-rank test and descriptive statistics were used to analyze the data.

\subsubsection{Results}
\begin{figure*}[hptb]
\centering
\includegraphics[width=0.9\linewidth]{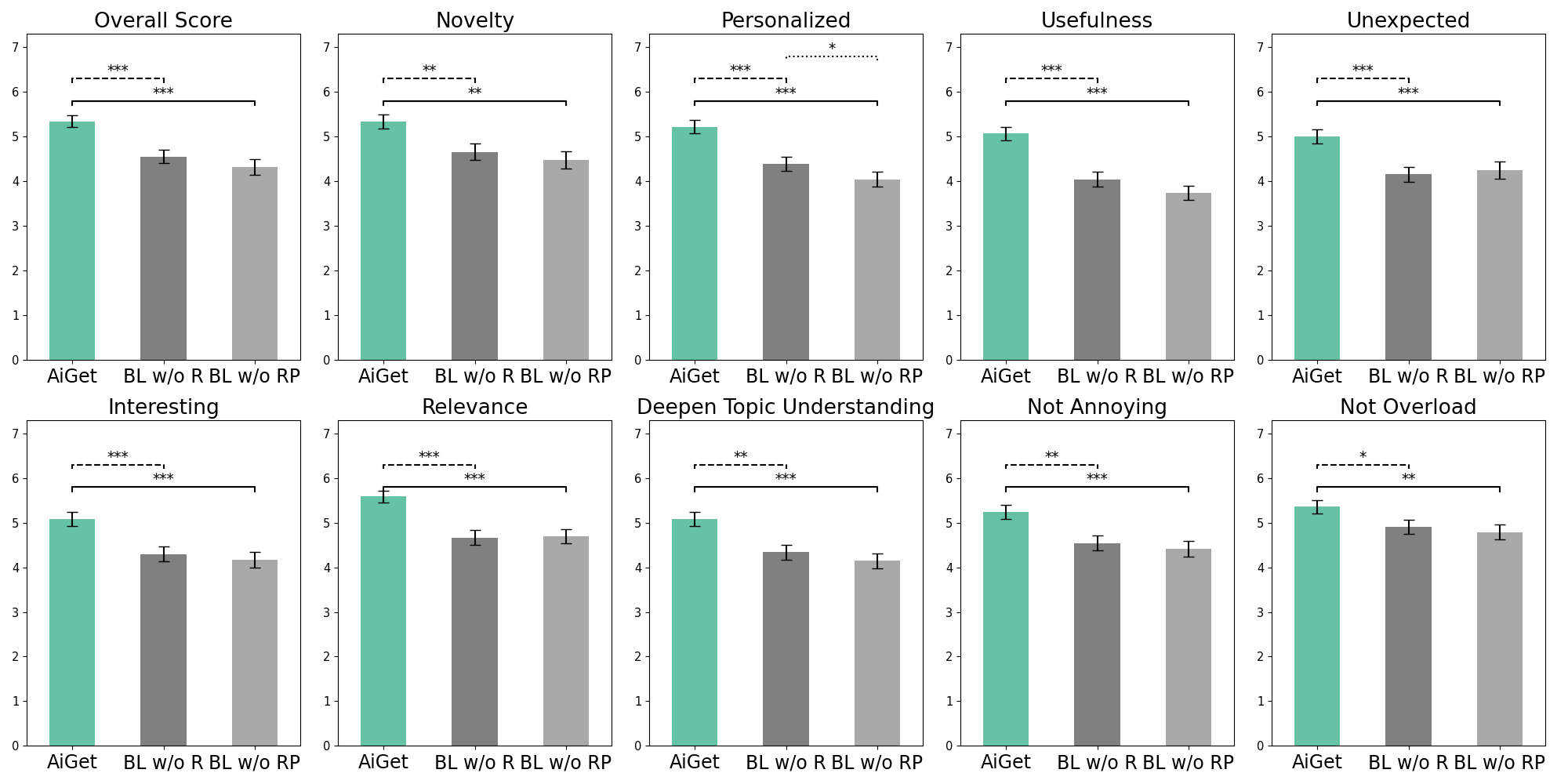}
\caption{Subjective ratings evaluating the desirability of generated knowledge. * indicates significance of \pval{<0.05}, ** indicates significance of \pval{<0.01}, *** indicates significance of \pval{<0.001}.}
\Description{This figure shows the subjective ratings evaluating the desirability of knowledge generated by AiGet, compared to two baseline conditions (BL w/o R: Baseline without Recommendations, and BL w/o RP: Baseline without Recommendations and Personalization). The ratings are presented across multiple dimensions: Overall Score, Novelty, Personalized, Usefulness, Unexpected, Interesting, Relevance, Deepen Topic Understanding, Not Annoying, and Not Overload. For each dimension, AiGet consistently scores higher than both baseline conditions, and statistically significant differences are indicated between the systems. Stars on the graph denote statistical significance levels, where p < 0.05 is indicated by *, ** indicates p < 0.01, and p < 0.001 is indicated by ***. AiGet demonstrates significant improvements across all dimensions, particularly in the dimensions of Overall Score, Interesting, Relevance, Usefulness, and Unexpected, showing a strong preference for AiGet compared to the baseline systems. Error bars represent the standard error of the mean.}
\label{fig:study2:tech_eval_scores}
\end{figure*}

As shown in Figure~\ref{fig:study2:tech_eval_scores}, \AiGet{} pipeline achieved significantly higher scores (\pval{<0.05}) across all measures compared to the two baselines, with all scores above 5 out of 7. This demonstrates its ability to generate novel, personalized knowledge with serendipity while keeping the information non-annoying and non-overloading. There was no significant difference between the two baselines except in \personalization{} (\pval{<0.05}).

\paragraph{Impact of User Profile Alone: Comparing \BaselinewoR{} vs. \BaselinewoRP{}}

Surprisingly, while adding a simple user profile can enhance \personalization{}, a desired factor aligning with \designGoalPersonalization{}, \BaselinewoR{} (i.e., Multimodal LLM + Profile) did not significantly improve results like \overallScore{} compared to \BaselinewoRP{} (i.e., Naive Multimodal LLM), indicating that merely adding a basic user profile to a naive MLLM pipeline is insufficient to generate desired knowledge. This is because the naive MLLM pipeline with personalization produced irrelevant knowledge by overemphasizing "user interests" without considering the user's intentions during primary tasks. As shown in Figure~\ref{fig:study2:tech_eval_doc}, when P7, who indicated an interest in ``Japanese culture/language'' and ``alcohol'' in their user profile, focused on Japanese alcohol, \BaselinewoR{} introduced ``Japanese drinking etiquette'' instead of providing information about the alcohol itself, which was less satisfying than \BaselinewoRP{}'s basic but relevant alcohol category information.

This underscores the need for AI systems to dynamically infer user interests and intentions from real-time context, as users may not explicitly define all their fine-grained desires in advance. It also highlights the importance of coordinating rules with user profiles.

\paragraph{Impact of Knowledge Generation Rules with User Profile: Comparing \AiGet{} vs. \BaselinewoR{}}

\AiGet{}, combining knowledge generation rules with user profiles, provided significantly more desirable knowledge than \BaselinewoR{} (i.e., Multimodal LLM + Profile) by balancing personalization, context relevance, and alignment with the user's in-situ intentions, highlighting the importance of \designGoalPrioritization{}. In the example shown in Figure~\ref{fig:study2:tech_eval_doc}, \AiGet{} received the highest scores by correctly predicting the user's intention as "deciding which alcohol to purchase, possibly comparing brands," and identifying the user's \focused{} mode. Following the formative study's rules (Sec~\ref{sec:study1:results:rq3}) aligning with design goals (\designGoalPrioritization{}), it provided balanced information between user interest (e.g., the alcohol name's meaning in Japanese) and usefulness (e.g., comparative flavor profiles with nearby related options).

Additionally, \AiGet{} leveraged gaze pattern analysis to introduce unexpected knowledge about unseen objects at appropriate moments, increasing its utility. For instance, when analyzing P9's \saccade{} pattern during a casual walk, it introduced \quote{surprising} information about nearby but ``unseen'' white chickens. In contrast, when P5 \focused{} on skincare products, \AiGet{} delivered relevant product knowledge, while the baselines introduced irrelevant details (e.g., an air conditioner on the ceiling). Furthermore, \AiGet{}'s scoring/weighting and filtering mechanism helped avoid user annoyance by withholding undesired content in less interesting indoor areas for two moments.

This ability to balance personalization, usefulness, unexpectedness, and the timely withholding of uninterested content demonstrates \AiGet{}'s capability in tailoring information delivery---a capability the baseline systems could not achieve. 
Comparing all results, this highlights that while \designGoalPrioritization{} contributes more to overall knowledge desirability than \designGoalPersonalization{}, all \designGoalContextUnderstanding{}-\designGoalPrioritization{} guidelines are still compulsory.

\section{Using \AiGet{} in Real-World Scenarios}
\label{sec:study3}

In-lab evaluations demonstrated that the \AiGet{} pipeline generated knowledge with high levels of novelty, personalization, surprise, and usefulness. While these findings emphasize the system's potential to produce knowledge about its surroundings, its performance in real-time, real-world scenarios remains untested. Therefore, to assess the feasibility of \AiGet{} in everyday environments, we conducted a user study to address the following question:

\begin{itemize}
    \item \textbf{RQ1:} How feasible is \AiGet{} in creating learning opportunities across various scenarios?
    \item \textbf{RQ2:} How does \AiGet{} influence users' behaviors and daily routines?
    \item \textbf{RQ3:} How do users perceive the \AiGet{} experience with continued usage?
\end{itemize}

It is important to note that fully addressing these questions with a high degree of quantitative evidence would require a larger user base and longer periods of daily usage, which was beyond the scope of this study. However, our study sought to gather user reflections that could provide meaningful insights into these research questions.

Additionally, we did not directly compare the two modes of \AiGet{}--- AI-initiated proactive knowledge discovery and user-initiated VQA tasks---in a head-to-head manner. Instead of replacing user-initiated queries, our goal was to explore whether AI-initiated interactions could overcome some limitations of user-initiated queries and enhance informal learning about the surrounding environment. Therefore, both modes were available in this study, allowing users to choose freely, following methods from prior research \cite{outlinespark24wang}.

\subsection{Participants}
The study consisted of two phases. In \PhaseI{}, 18 participants (8 females, 10 males, Age: \meansd{23.9}{3.7}) from the university community engaged in a \textbf{one-time use} of \AiGet{}. Among these participants, six reported using AI tools (e.g., Google Lens, ChatGPT app) to identify surrounding objects in their daily activities at least three times per week. Eleven participants used such tools less frequently, ranging from once per month to twice per week. The remaining participant rarely used these tools.
All participants self-reported a minimum level of professional working fluency in English.

In \PhaseII{}, 6 participants (3 females, 3 males) from \PhaseI{} were selected for \textbf{continued usage} based on their demographic diversity and availability (see Appendix~\ref{appendix:real_world_study:demographics} for details). These participants, representing a range of self-reported engagement with their surroundings---3 with high engagement and 3 with lower engagement---took part in additional experiments involving multiple sessions.

\subsection{Apparatus}
\label{sec:study3:apparatus}
As depicted in Figure~\ref{fig:teaser} and detailed in Sec~\ref{sec:system:implementation}, participants wore smart glasses and a ring mouse as part of the \AiGet{} system. Participants also carried a lightweight backpack to house the computing unit (i.e., the laptop) during sessions. 

The experimenter used a mobile phone (iPhone 12) to record the participants' behaviors and provided a hotspot for internet connectivity.

\begin{figure}[h]
\centering
\includegraphics[width=1\linewidth]{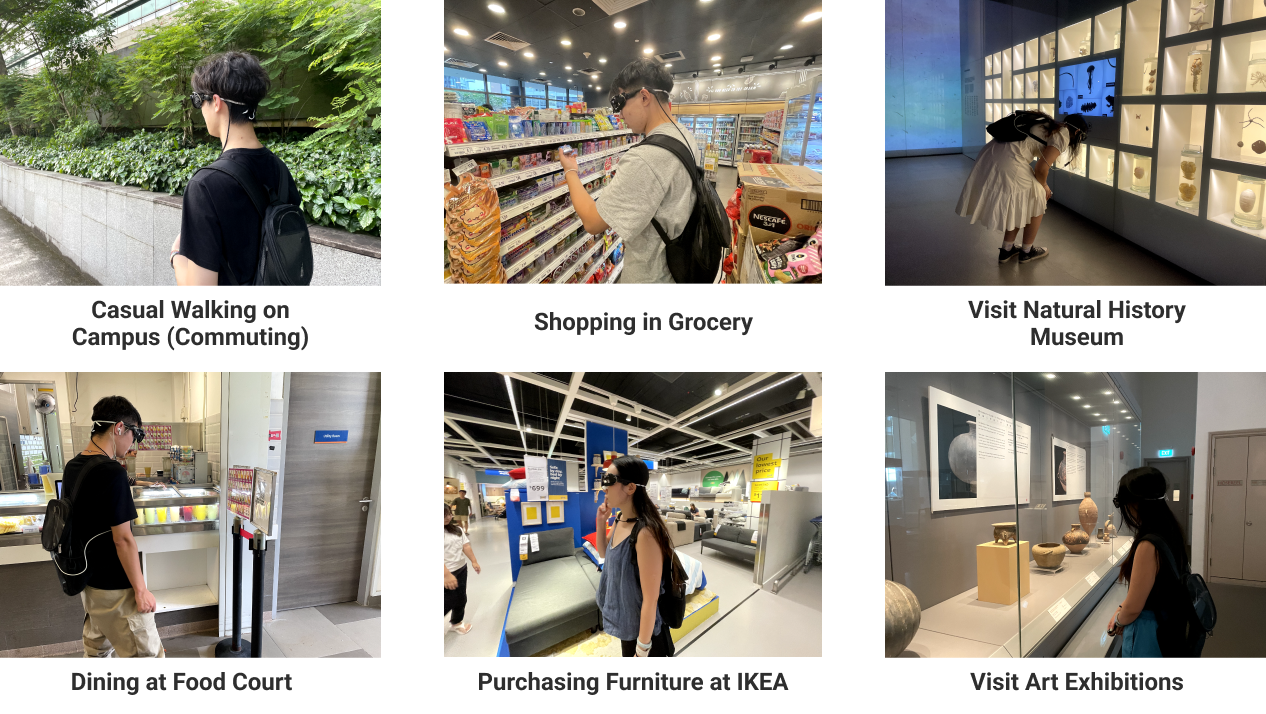}
\caption{Different Usage Scenarios with \AiGet{}.}
\Description{This figure shows various usage scenarios with the AiGet System, including commuting, dining, shopping, and visiting natural history museums and art exhibitions.}
\label{fig:study3:usage_scenarios}
\end{figure}

\subsection{Study Design}

In \PhaseI{}, 18 participants were randomly assigned to one of three settings: casual walking on campus, shopping, or visiting a local natural history museum, with 6 participants in each setting. These settings were chosen to represent high, medium, and low frequencies of daily activities, respectively, but with potentially increasing levels of information acquisition desire in the same order. 

In \PhaseII{}, 6 participants completed at least two additional sessions over the following 7 days, resulting in a minimum of 3 sessions per participant. This aligns with prior research on using wearable devices for informal learning in daily routines \cite{audioxtend24tan}. Although participants could not take the device home due to the limited availability of only one prototype, they were able to schedule sessions at any preferred time and location with the experimenters. This led to the exploration of various scenarios, including dining at the canteen, window shopping at a mall, purchasing furniture at IKEA, and visiting art exhibitions, as shown in Figure~\ref{fig:study3:usage_scenarios}. To assess the system's continued/repeated usage in the same places, participants were required to revisit at least one location.

In total, 18 sessions were completed in \PhaseI{}, and 6 out of the 18 participants conducted an additional 22 sessions (P1: 4, P2: 6, P3: 6, P4: 2, P5: 2, P6: 2 sessions) in \PhaseII{}, resulting in 40 sessions in total.

\subsection{Procedure}
In \PhaseI{}, participants signed a consent form, received training, and used the system for at least 45 minutes at designated places. In \PhaseII{}, participants were encouraged to use the system for at least 30 minutes per session at their desired locations. After each session in both \PhaseI{} and \PhaseII{}, participants completed a questionnaire and participated in a 15-minute interview to share their experiences.

\subsection{Measures}
To evaluate the feasibility of \AiGet{} for in-the-wild informal learning, we gathered data on the following key measures: 1) the desirability of the knowledge provided by the system, 2) the usability of the system while engaged in primary tasks, and 3) the system's interference with participants' ongoing primary tasks. For details, see Appendix~\ref{appendix:measures}.

\subsubsection{Desirability of Knowledge Acquisition} 
To assess the desirability of knowledge provided by \AiGet{}, we employed both subjective and objective measures. Subjective measures included all those from the in-lab evaluation (Sec~\ref{sec:study2:measures}), along with a new measure, \deepenEnvironment{} (7-point Likert scale), to determine whether \AiGet{} enhances users' connection with their surrounding environment.

Objective measures included tracking the count of AI-initiated knowledge, the frequency of canceled displays, user-initiated queries \cite{pandalens24cai}, and the immediate recall of novel knowledge. Additionally, we recorded instances where participants perceived the system as providing incorrect information.

\subsubsection{Usability of the System with Primary Tasks}
We measured usability using the System Usability Scale (\SUS{}) \cite{brooke_sus_1996} and assessed perceived cognitive load through the \perceivedTaskLoad{} scale \cite{nasa_tlx_2006}, providing insights into the effort required to use \AiGet{} while managing primary tasks.

\subsubsection{Interference with Primary Tasks} 
Following prior research on wearable assistants and interactions with smart glasses during daily activities \cite{pandalens24cai, cai2023paraglassmenu, janaka_paracentral_2022}, we assessed participants' experiences with their primary tasks, including \distraction{}, \enjoyment{}, and \naturalness{}, using 7-point Likert scales.

\subsection{Findings}
\label{sec:study3:findings}

Table~\ref{tab:summary_stats_category} and Figure~\ref{fig:study3:subjective_different_scenarios} present the descriptive statistics for all measures. Following the same analysis procedure as in the formative study (Sec~\ref{sec:study1:analysis}), we address our research questions.

\subsubsection{RQ1: How feasible is \AiGet{} in creating learning opportunities across various scenarios?} 
\label{sec:study3:results:rq1}

Before conducting the user study, we had several concerns about how users might perceive \AiGet{}. These concerns were focused on three main points: (1) uncertainty about the quality and relevance of AI-generated knowledge in real-world testing, (2) whether information delivered proactively would disrupt or annoy users during their daily activities, and (3) doubts about how valuable users would find information they didn't ask for. These concerns reflected broader questions about how to effectively integrate AI-driven informal learning into daily life. We were particularly interested in seeing how users would react to and benefit from this new approach to learning and information delivery.

\begin{figure*}[hptb]
\centering
\includegraphics[width=1\linewidth]{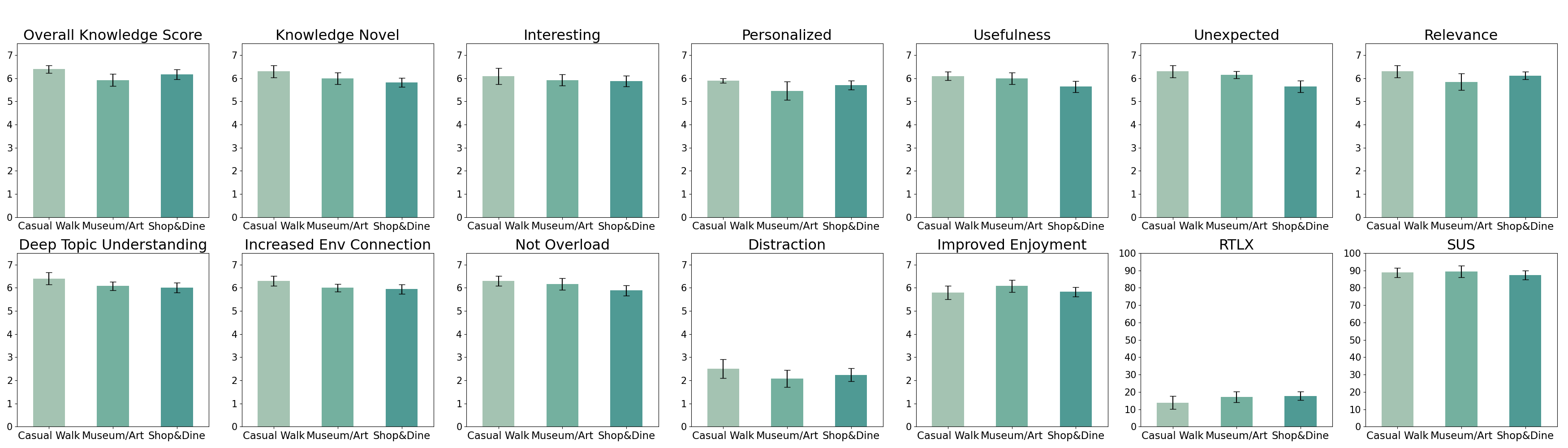}
\caption{Subjective ratings on the desirability of generated knowledge, its impact on primary tasks, and system usability in different scenarios.}
\Description{This figure presents subjective ratings on the desirability of the generated knowledge, its impact on primary tasks, and system usability across three different scenarios: Casual Walk, Museum/Art, and Shop\&Dine. The ratings are measured across several dimensions: Overall Knowledge Score, Knowledge Novel, Interesting, Personalized, Usefulness, Unexpected, Relevance, Deep Topic Understanding, Deep Environment Connection, Not Overload, Distraction, Improved Enjoyment, RTLX (NASA Task Load Index), and SUS (System Usability Scale). Overall, the results show strong similarities in user ratings across the different scenarios. Metrics like Overall Knowledge Score, Knowledge Novel, Interesting, Personalized, Unexpected, Deep Topic Understanding, and Improved Enjoyment are consistently high across all scenarios, indicating a uniform positive perception of the generated knowledge. Distraction ratings remain low across the board, suggesting that users did not find the system intrusive during primary tasks. Additionally, Not Overload and Deep Environment Connection maintain stable, high scores, reflecting the system’s ability to integrate knowledge without overwhelming the user. RTLX scores are similarly low across scenarios, indicating minimal cognitive load. SUS scores, representing system usability, are consistently high across all scenarios, reflecting the system's ease of use.}
\label{fig:study3:subjective_different_scenarios}
\end{figure*}

The results were notably positive, surpassing our initial expectations and alleviating our concerns. Participants spent an average of 43.0 minutes (SD = 13.6) per session using \AiGet{}. Addressing our first concern, they found \AiGet{} provided useful (5.9 out of 7, on average), interesting (6.0 out of 7),  surprisingly unexpected (6.0 out of 7), and relevant (6.1 out of 7) knowledge, deepening their understanding on the topics (6.1 out of 7). Contrary to our second concern, while proactively prompting users with 1.26 knowledge per minute on average, \AiGet{} did not cause significant distraction (2.2 out of 7) to users' primary tasks. Resolving our third concern, it helped all participants gain "unseen and unknown knowledge" in various daily activities, making them feel like they had a \quoteby{P4}{friend} or \quoteby{P2, P6, P11}{companion}, and fostered a deep connection with the environment (6.0 out of 7). Additionally, participants perceived \AiGet{} as having `Excellent' usability (\SUS{}: 88.4, higher than 80 \cite{bangor_empirical_2008}), while incurring a low cognitive load (\perceivedTaskLoad{}: 16.6) and perceiving natural (6.0 out of 7) to use during primary tasks, further supporting its successful integration into everyday life.

As expected, \AiGet{}'s proactive AI-initiation generated significantly more knowledge (M=54.2) compared to passive User-initiation (M=2.3). Notably, despite this substantial increase in knowledge delivery, AI-initiation had a surprisingly low cancellation rate (2.4/54.2 $\approx$ 4\%), indicating that users found proactive knowledge delivery acceptable most of the time. Additionally, users recalled an average of 7.8 novel knowledge instances per session (AI-initiation: M=7.1, User-initiation: M=0.7), highlighting \AiGet{}'s knowledge acquisition support.

Together, these findings underscore \AiGet{}'s ability to help users acquire more knowledge than through self-directed queries alone, overcoming barriers like \quoteby{P16}{I don't know what to ask}. Furthermore, participants acknowledged that both AI-initiated and user-initiated interactions play \quoteby{P7}{supplementary roles}. For instance, P4 shared that \quote{initially, I have no question as I don't know what it [flying squirrel] is until it [\AiGet{}] shares information with me, so I asked a follow-up question with this inspiration.} This demonstrates how proactive `desirable' knowledge delivery can spark curiosity and encourage further exploration by users.

\begin{table*}[hptb]
\centering
\resizebox{\textwidth}{!}{%
\begin{tabular}{lcccccc}
\toprule
Scenario (Count) & Duration (min) & AI-Initiated Knowledge & Cancel Display & Immediate Recall & User-Initiated Query & Perceived Wrong \\
\midrule
Casual Walking (10) & \meansd{43.6}{17.6} & \meansd{51.1}{24.8} & \meansd{2.8}{2.6} & \meansd{6.8}{3.9} & \meansd{1.7}{2.5} & \meansd{0.1}{0.3} \\
Museum \& Art Exhibition (13) & \meansd{46.8}{13.5} & \meansd{62.4}{20.0} & \meansd{2.4}{2.8} & \meansd{7.9}{5.6} & \meansd{3.0}{2.5} & \meansd{1.6}{1.0} \\
Shopping \& Dining (17) & \meansd{39.6}{10.8} & \meansd{49.9}{15.7} & \meansd{1.9}{3.3} & \meansd{8.2}{5.1} & \meansd{2.4}{3.0} & \meansd{0.8}{0.9} \\
\midrule
All Sessions (40) & \meansd{43.0}{13.6} & \meansd{54.2}{20.0} & \meansd{2.4}{2.9} & \meansd{7.8}{4.9} & \meansd{2.3}{2.9} & \meansd{0.8}{1.0} \\
\bottomrule
\end{tabular}%
}
\caption{Objective descriptive statistics for \AiGet{} system usage in different scenarios.}
\Description{This table provides objective descriptive statistics of AiGet system usage across three different scenarios: Casual Walking (10 sessions), Museum \& Art Exhibition (13 sessions), and Shopping \& Dining (17 sessions). It includes data on session Duration (in minutes), the number of AI-Initiated Knowledge prompts, the number of Cancel Display actions, the number of Immediate Recall instances, the number of User-Initiated Queries, and Perceived Wrong outputs.}
\label{tab:summary_stats_category}
\end{table*}

\paragraph{Usage in Different Scenarios}
\label{study3:results:rq1:different_scenarios}
As shown in Figure~\ref{fig:study3:subjective_different_scenarios},  \AiGet{} provides an equally good experience in acquiring knowledge with enhanced enjoyment for primary tasks in different scenarios. 
Interestingly, our study revealed that users perceived \AiGet{} in diverse roles across different contexts, despite its consistent knowledge-generation mechanism. These roles emerged naturally, shaped by users' expectations, goals, and activity contexts. This highlights how users' mindsets and situational needs influence their interpretation of \AiGet{}, affecting how they integrate and value the information in various aspects during daily activities.

\begin{figure}[hptb]
\centering
\includegraphics[width=1\linewidth]{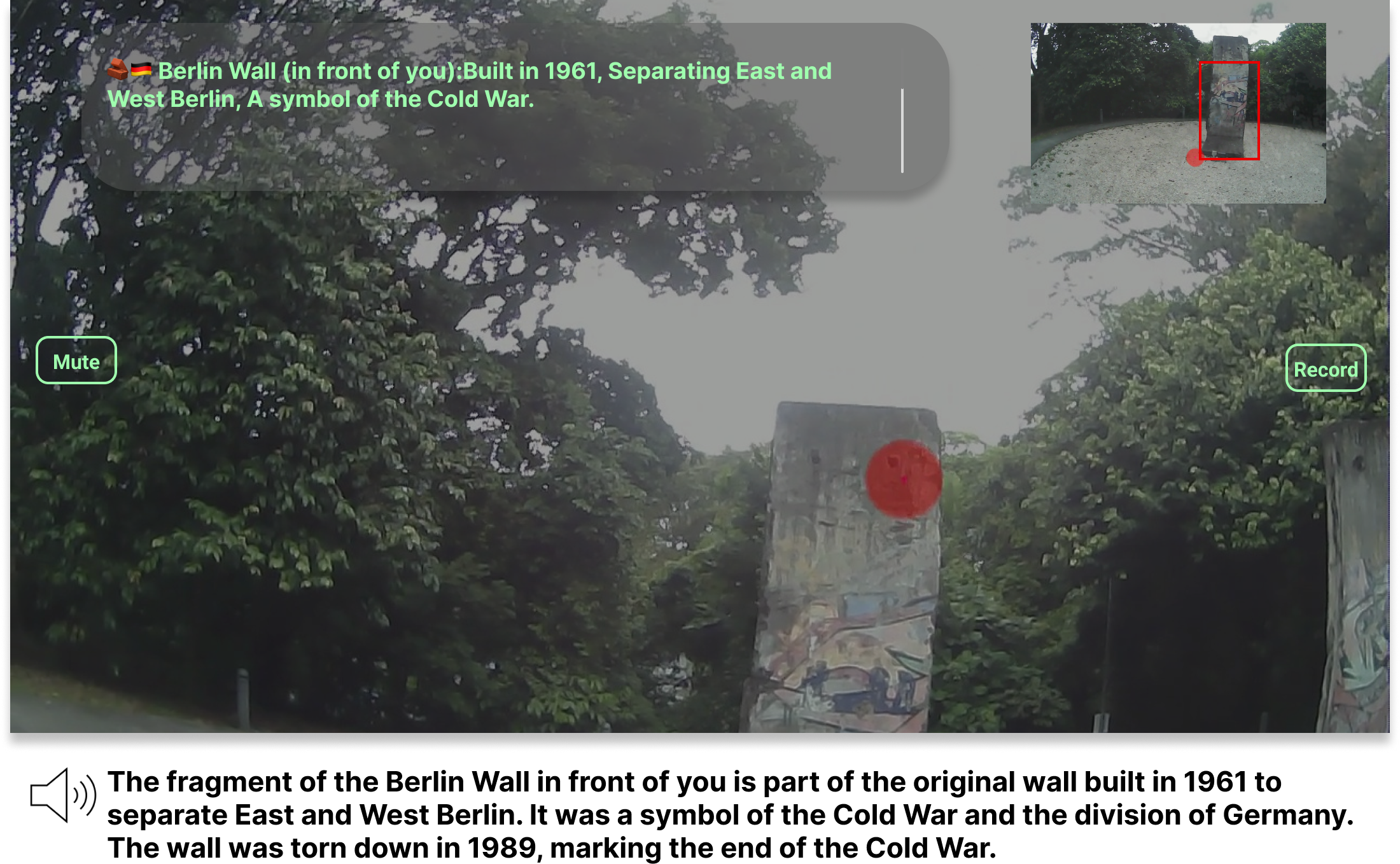}
\caption{An example of \AiGet{} helping users identify `unknown unknowns' in daily life: two walls on campus are part of the Berlin Wall, with accompanying information introducing its history.}
\Description{This figure shows an example of the AiGet System helping users realize that two walls on campus are part of the Berlin Wall and introduces its history.}
\label{fig:study3:berlin_wall}
\end{figure}

(1) During \textit{Casual Walking in Familiar Places}, users perceived \AiGet{} as \textbf{``Environmental Storyteller''} to introduce \quote{Familiar Strangers} in their environment. This included knowledge that interested users but was unexplored due to time constraints (i.e., \textbf{known unknowns}) and knowledge users hadn't anticipated (i.e., \textbf{unknown unknowns}) about the environment. For example,
P2 appreciated when \AiGet{} detected the school logo and explained the meaning behind its colors, \quote{I have a design background and knew the general color meaning, but I had never connected them to the school logo until \AiGet{} told me. It was a \textbf{pleasant surprise} that strengthened my connection to the school.} Additionally, as demonstrated in Figure~\ref{fig:study3:berlin_wall}, 
P9 was amazed to learn that two concrete walls on campus were actually Berlin Wall, \quote{I frequently passed by but never expected to find pieces of the real Berlin Wall on our campus.}
This type of discovery was particularly well-received during casual walks, where users were in a more relaxed mindset, making them more receptive to narrative-like information about their surroundings.

(2) During \textit{Shopping \& Dining},  \AiGet{} was seen as a \textbf{``Personalized Decision-Making Supporter.''} 
It helped \textbf{mitigate uncertainty avoidance} ~\cite{money2003uncertainty_avoid} and facilitate informed choices by providing customized introductions, breaking down language barriers, and aligning suggestions with personal preferences.
For example, P1 and P4 had previously avoided imported beverages and food with foreign language labels due to unfamiliarity and language barriers. However, with \AiGet{}, P1 found he could easily \quote{navigate around and understand what I'm looking at,} ultimately selecting a suitable drink he had never tried before. Similarly, P2 ordered a dish he had never tried until \AiGet{} introduced its unique cooking method and taste, aligning with P2's preferences. Furthermore, \AiGet{} helped P6 \textbf{make quick choices and notice unseen but useful options}, aligning with her health values:
\quote{It saved me time by telling me which bean was lower in calories, comparing them with similar ones that I hadn't noticed on the same shelf.} Thus, these highlighted benefits of perceived role align with users' primary goal in these contexts: making decisions.

(3) In \textit{Museum \& Art Exhibitions}, \AiGet{}'s role was perceived as a \textbf{``Personalized Information Highlighter''}. It enhanced the educational experience by assisting users in extracting key information from exhibition boards, helped them notice themes or elements they might have overlooked, and provided additional facts aligned with their personal interests.
P3, for example, often scans museum boards quickly, missing certain details. \AiGet{}, aware of her design background, highlighted how metal elements affect ceramic colors, which she initially overlooked. It also provided surprising information about the Persian origin of cobalt blue dyes, expanding her knowledge.
Similarly, P14 appreciated how \AiGet{} highlighted the constant replacement of shark teeth, something she hadn't noticed during her initial scan, enhancing her learning in a concise and efficient manner. This perception likely arises from users' goal of absorbing information efficiently in potentially overwhelming educational settings.

These cases revealed that, despite using the same information generation mechanism, \AiGet{} becomes naturally useful in diverse contexts by tailoring information to a person's preferences and background. This adaptability highlights the potential of AI systems to seamlessly integrate into various aspects of daily life. It suggests that well-designed AI assistants can enhance user experiences across different scenarios without altering their core mechanisms. Instead, their perceived value and functionality evolve based on the user's needs and context, opening new possibilities for designing versatile AI systems that fluidly adapt to changing circumstances and goals.

\subsubsection{RQ2: How does \AiGet{} influence users' behaviors and daily routines?\\} 
\label{sec:study3:results:rq2}

While our initial goal for \AiGet{} was modest---helping users become aware of initial ignored yet desirable knowledge from everyday moments---the study results exceeded our expectations. We discovered that \AiGet{} is more than just a tool for displaying hidden knowledge; it shows potential as a medium that shapes user behavior and transforms daily routines, drawing users' \quoteby{P1}{attention back to their in-situ environment}. This shift has three main effects.

(1) Most (16/18) users reported it made them \quoteby{P13}{more observant and increased their curiosity about their surroundings}. This shift addresses users' previous \quoteby{P1}{short attention span due to smartphone usage} and task-driven lifestyle, which often led to behaviors such as \quoteby{P2}{rushing during commuting}, \quoteby{P16}{skimming in museums}, and \quoteby{P17}{shopping with only specific items in mind}. Now, users are more interested in their surrounding environment and willing to explore further, as they can obtain useful and surprising knowledge from the system. By \quoteby{P15}{fostering curiosity} and \quoteby{P7}{deeper engagement}, \AiGet{} helps users shift from passive interaction with their surroundings to a more exploratory mindset, unlocking \quote{hidden layers [e.g., perspectives, benefits]} of their everyday environments.

(2) One such unlocked \quote{hidden layers} brought by prolonged attention is to revive suspended personal interests or hidden wishes. For instance, P8, a biology major with an interest in typography, was surprised when \AiGet{} provided an unexpected perspective: \quoteby{P8}{I usually enjoy looking at posters and notices, and \AiGet{} surprised me with information about font design [Helvetica] on a board. It was quite pleasant.} Similarly, P15, who had a long-term hidden interest in car models, now enjoys walking by the road with \AiGet{}, gaining deeper knowledge about each model's characteristics. This demonstrates how AI systems like \AiGet{} can bring unexpected joy and discovery, aligning with personal interests into routine activities, rekindling forgotten passions, and creating richer experiences in users' daily lives.

(3) Building on these benefits, participants noted \AiGet{}'s potential to reduce reliance on smartphones for social media in some situations, a primary method for informal learning as mentioned in the formative study (Sec~\ref{sec:study1:results:rq1}). While social media offers \quoteby{P14}{in-depth information on trending topics} and is \quoteby{P6}{suitable for use at home}, participants highlighted \AiGet{}'s ability to replace social media during \quoteby{P6}{everyday activities}, providing a \quoteby{P14}{broader understanding of the world}. Three participants reported less phone usage while queuing for checkout in stores: \quoteby{P2}{Previously, I would check social media to pass time. Now, I was observing my surroundings and checking products because I could gain interesting information related to my current environment.}
\AiGet{} was also perceived as more memorable and connected to the environment than social media. P9 commented: \quote{It gives me new information about the environment that I will remember and explore further, but information from my phone is often forgotten after I see it.} P15 added: \quote{If I'm scrolling through TikTok, I feel it's less useful. But this [\AiGet{}] helps you understand your environment, which could be useful later on.}

The above cases indicate that, while changing habits in daily routines is typically challenging, \AiGet{} offers a unique approach by delivering engaging and valuable information that enables users to naturally and effortlessly adjust their behaviors. This points to the potential for future AI assistants to subtly and positively influence user habits, making behavior change seamless and enjoyable through meaningful experiences.

\subsubsection{RQ3: How do users perceive the \AiGet{} experience with continued usage?}
\label{sec:study3:results:rq3}

\begin{figure*}[h]
\centering
\includegraphics[width=1\linewidth]{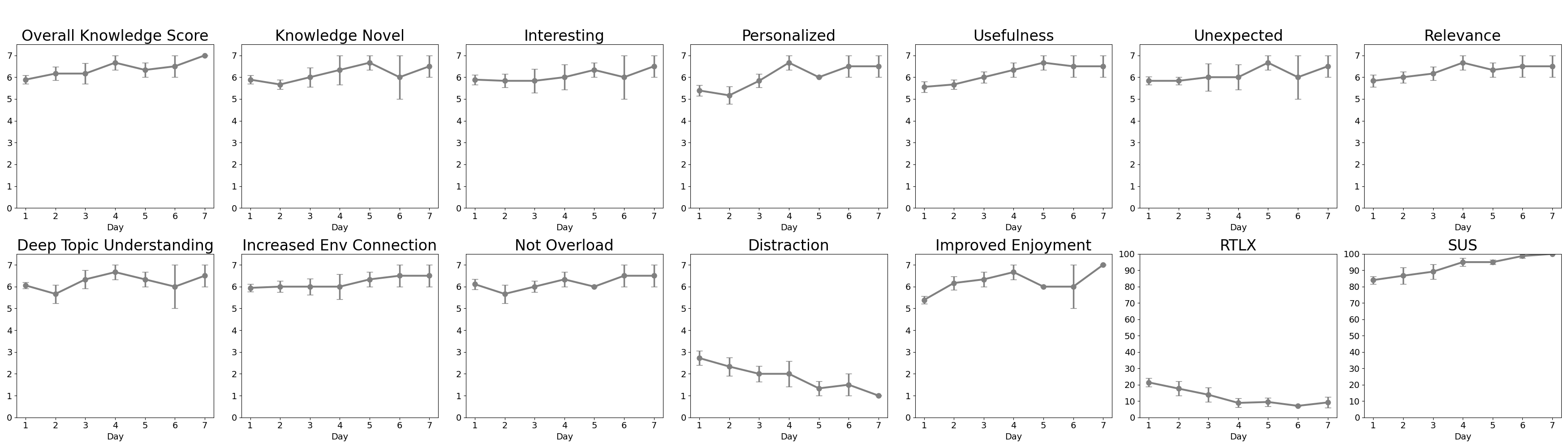}
\caption{Subjective ratings on the desirability of generated knowledge, its impact on primary tasks, and system usability on different days (up to 7 days).}
\Description{This figure presents the subjective ratings of the desirability of generated knowledge, its impact on primary tasks, and system usability over seven days. The metrics displayed include Overall Knowledge Score, Knowledge Novel, Interesting, Personalized, Usefulness, Unexpected, Relevance, Deep Topic Understanding, Deep Environment Connection, Not Overload, Distraction, Improved Enjoyment, RTLX, and SUS. Across the days, most of the metrics have slight increases. Distraction shows a decreasing trend over the seven days, suggesting that the system becomes less distracting with repeated use. Similarly, RTLX, a measure of cognitive load, decreases consistently, implying reduced task load as users become more accustomed to the system.}
\label{fig:study3:subjective_different_trials}
\end{figure*}

While our study was conducted up to 7 sessions per participant across different days, insufficient for long-term usage analysis, we gathered initial feedback on continued usage in various locations.
As illustrated in Figure~\ref{fig:study3:subjective_different_trials}, the positive experience of consistently gaining desired knowledge, with minimal interference to primary tasks, did not diminish after continued usage. Instead, it demonstrated an increasing trend. We further examined user feedback after multiple usage in different or the same places.

\paragraph{Usage in Different Places (16 out of 22 additional usages)} 
Most (5/6) participants appreciated usage at different places as they provided different values, as also demonstrated in RQ1 (Sec~\ref{study3:results:rq1:different_scenarios}). Interestingly, over time, participants developed personalized usage habits, leveraging the system's flexible output options to match their knowledge acquisition needs and mindsets in various scenarios. For example, while shopping, P4 preferred to mute the system and receive text-only feedback for quick, information-dense updates, which enables him to quickly \quote{browse various products in a short time and glance at the information only when needed}. But during casual walks, he opted for audio feedback to relax.

However, one participant, P5, preferred limiting the system's use to specific contexts (e.g., shopping). Valuing \quote{efficiency of acquiring information in life}, she found that knowledge that is merely \quote{good to have} (e.g., surprise by the bird in the environment) wasn't worth the cognitive effort if it lacked immediate usefulness.

These cases emphasize the importance of designing proactive AI systems with flexible output options and on/off controls, catering to different user preferences and the varying demands of long-term usage in diverse scenarios.

\paragraph{Usage in Same Places (6 out of 22 additional usages)} 
With a design to filter similar knowledge and provide alternative perspectives of the same entities (Sec~\ref{sec:system:knowledge_generation}), repeated exposure to the same place did not reduce novel knowledge intake (First Time vs. Second Time \novelty{}: 5.7 vs. 5.8 out of 7) as long as the environment was \quoteby{P6}{dynamic and large}. As P1 described it, \quoteby{P1}{it's like hunting new things that I never found.}

Interestingly, because participants could become more observant, as mentioned in Sec~\ref{sec:study3:results:rq2}, it helped them find new interests in familiar environments, even 
when the \AiGet{} system was not in use.
P2, who used \AiGet{} for 7 days, mentioned that before using it, he usually didn't pay attention to his familiar environment during his commute. However, after using the system (and twice in commuting), he became more observant, \textbf{even without} \AiGet{}, and noticed a Hornbill building a nest one morning by a school building, wishing he had the system to learn more about Hornbills.

Although long-term usage still needs further investigation, these use cases highlight the feasibility and utility of proactive AI assistants for informal learning.

\section{Overall Discussion}

The \AiGet{} system demonstrates a potential for transforming everyday moments into informal learning opportunities by effectively addressing longstanding barriers to learning from the in-situ environment. By proactively analyzing user attention and context, \AiGet{} uncovers knowledge often overlooked, bridging the gap from "unseen" to "seen", from "seen" to "known", and from "known" to "known better". This minimized-effort approach deepens user engagement with their surroundings, enhancing participants' curiosity and observational skills while reviving hidden interests and fostering a willingness to explore both familiar and unfamiliar environments.

In the following discussion, we will highlight two key aspects: 1) how wearable knowledge discovery assistants (i.e., wearable AI assistants supporting knowledge discovery) can be used for informal learning in daily life, and 2) what improvements are needed for their long-term usage. By examining these aspects, we aim to provide insights into the future development and integration of context-aware wearable knowledge discovery assistants in everyday contexts, further improving how we acquire and interact with information from our in-situ environment.

\subsection{How Wearable Knowledge Discovery Assistants Can Be Used for Informal Learning in Daily Life?}

\subsubsection{Effective Components for Addressing Cognitive Limitations and Biases}
\label{discussion:biases}

As shown in the real-world study (Sec~\ref{sec:study3:results:rq1}), the \AiGet{} system demonstrates the ability to reduce cognitive limitations and biases associated with informal learning through the following three key components.

\textit{Proactivity with Context Awareness.} 
This feature addresses "Known Unknowns" arising from time constraints, attentional limitations, and uncertainty avoidance \cite{baron2023thinking, money2003uncertainty_avoid} (Sec~\ref{sec:study1:results:rq1}) by proactively delivering context-relevant information with reduced manual effort. This approach has encouraged users to revive hidden interests and leave their comfort zones, leading to \quoteby{P4}{increased interaction with unfamiliar environments}. Moreover, such proactivity helps correct \quote{wrong assumptions} users often make when encountering unfamiliar entities (Sec~\ref{sec:study1:results:rq1}). For example, in the real-world study, P2 saw a vending machine on campus and assumed it was selling ice cream based on the packaging and refrigeration. \AiGet{} informed him it was actually selling protein bars, surprising P2 and making him remember the machine for \quote{future use before workouts}.

\textit{Gaze Pattern Analysis.} 
By analyzing user gaze patterns and identifying information in \textbf{both} gazed and ignored (i.e., not gazed) areas of the in-situ environment, \AiGet{} addresses inattentional blindness \cite{mack2003inattentional} (Sec~\ref{sec:study1:results:rq1}). This helps identify unseen yet interesting or useful information that users might otherwise overlook.

\textit{Extensive Knowledge Access with Personalization.}
By utilizing the vast knowledge within LLMs (derived from training data) and tailoring it to user familiarity through personalized profiles, \AiGet{} presents "Unknown Unknown" perspectives, often surprising users. This addresses the Illusion of Explanatory Depth \cite{rozenblit2002illusion_explanatory_depth} (Sec~\ref{sec:study1:results:rq1}) and deepens users' understanding of their surroundings, providing unexpected insights, especially for seemingly familiar objects or concepts.

Each of these components addresses distinct cognitive limitations and biases, highlighting the value of integrated systems like \AiGet{} in enhancing informal learning opportunities in everyday life.

\subsubsection{Potential Caveats of AI Biases and Design Considerations}
\label{discussion:biases:AI_Biases}
Although utilizing AI, especially LLM's extensive knowledge base, shows promise, careful design is crucial to avoid introducing AI biases. This is particularly important when presenting information about entities or topics with \textbf{both} positive and negative aspects. For instance, full-fat milk can be interpreted as either `nutritious due to its vitamins and minerals' or as `high in fat content'. For a health-conscious user, deciding which perspective to prioritize is crucial.
To address this challenge, the AI needs to understand more nuanced user preferences and contexts. For instance, if a user is ill, the system could emphasize the nutritional benefits of full-fat milk, whereas for a user focused on weight management, it could highlight its high-calorie content. This fine-tuning ensures the information aligns with the user's situational needs.

Additionally, the AI can present multiple perspectives over time, alternating between positive and negative aspects to provide a well-rounded understanding of complex topics. This approach helps prevent biased or one-sided information while tailoring the experience to users' evolving needs.

\subsubsection{Improving User Trust for ``Surprising'' Content}

Enhancing user trust in wearable knowledge discovery assistants is crucial, particularly when the system presents surprising or counterintuitive information. As P1 noted, prior negative experiences with AI-generated content (e.g., hallucinations \cite{bang2023hallucination}) can cause users to distrust unexpected information, even when it is accurate---such as learning that `a palm tree is more closely related to grass than typical trees'. 

To address this issue, future systems should incorporate several key features. First, a more refined scoring/prioritization system is needed. Rather than relying on the binary assessments used in the current \AiGet{} prototype (Sec~\ref{sec:system:knowledge_generation}), future versions could implement a nuanced scoring system to measure ``unexpectedness.'' When this score is high, the system could provide reference marks to enhance credibility, similar to practices in some existing AI systems (e.g., \href{https://www.perplexity.ai/}{perplexity.ai}). Additionally, a "check source" button could be introduced, allowing users to verify detailed and authentic explanations as needed. This design of offering references only when necessary enhances trust without overwhelming users, maintaining a seamless experience for consuming information on the go.

\subsection{Improvements for Long-Term Usage of Context-Aware Wearable Knowledge Discovery Assistants}
\label{sec:discussion:future_improvements}

\subsubsection{Balancing Knowledge Diversity and Depth}
In the real-world study, participants highlighted \AiGet{}'s ability to \quoteby{P3}{break out of information cocoons [exposed only to information that aligns with their existing beliefs or interests]} by providing information on both prior interested topics and relatively unfamiliar content embedded in the environment, thus expanding their \quoteby{P18}{knowledge base}. While this feature fosters the discovery of new interests, it also raises concerns about potential information overload. As P1 noted, \quote{I don't need to be an expert on everything,} emphasizing the need for careful design in content delivery, especially for repeated exposure to entities of limited interest.

To enhance long-term engagement with wearable knowledge discovery assistants, we propose a dynamic approach that balances knowledge diversity and depth, adjusting to user interests and environmental changes over time. Based on the knowledge participants gained in our studies, we categorize knowledge depth into three levels, aligned with the progression from "seen" to "known" to "known well":
(Level 1) Alerting users to the existence of unseen, new, or changed entities in the environment.
(Level 2) Providing surface-level yet interesting or surprising information about these "seen" entities.
(Level 3) Delivering detailed or expert-level information tailored to users' specific interests.

For repetitive use in relatively familiar environments, the system should prioritize Levels 1 and 2 when users are \saccade{-ing} or \quickBrowse{-ing} (Sec~\ref{sec:study1:results:rq2}). This approach helps users discover accessible knowledge and build broader connections with the environment. Level 3 should be reserved for topics where users have demonstrated sustained interest (e.g., focusing on something), fostering deeper connections with subjects they find interesting.
This tiered, context-adaptive approach balances breadth and depth, mitigating overload while personalizing the learning experience.

\subsubsection{Enhancing Coherence Between Learning Moments}
Participants suggested that future systems could strengthen learning by linking new information to previous experiences, supporting scaffolding \cite{valsiner_culture_1997} to build structured knowledge for long-term retention and creating more memorable interactions. This idea aligns with research on LLM companions \cite{xu2024can}, which use historical context to enhance user engagement and comprehension, and wearable memory assistants like Memoro \cite{zulfikar2024memoro}, which use past context to augment user memory.

For example, two participants mentioned that \AiGet{}'s knowledge prompts reminded them of their undergraduate studies, making the information more memorable. However, during P14's museum visit, \AiGet{} initially provided details on limb evolution from fins but missed the opportunity to connect this information to a Dugong\footnote{The dugong is a highly endangered marine mammal, which is commonly known as `sea cows'.} specimen with limb-like features later in the visit. This oversight disappointed the participant and highlighted the need for future systems to automatically link related information across different moments. To address this, future systems could use techniques like knowledge graphs or Retrieval-Augmented Generation (RAG) \cite{zamani_retrieval_2022} to identify and connect relevant knowledge across various encounters.

\subsubsection{Considering Interruptibility in Real-Time Mode and Supporting Retrospective Mode}
\label{discussion:long_term:async}
To reduce the interference to primary tasks, beyond providing easy-to-digest content and enabling manual output adjustment, the future system could further predict user interruptibility \cite{pejovic_interruptme_2014} and automatically reduce its frequency or turn off during moments when users prefer fewer interruptions (as shown in Figure~\ref{fig:study1:framework}-Gray parts). This feature can be implemented by 1) Using MLLMs for FPV analysis and withholding notifications when users are engaged in high-concentration or time-sensitive activities (e.g., crossing a road, focusing on work, and rushing in train stations); 2) Incorporating additional input modalities, such as Electrodermal Activity signals \cite{tan2021neuro_interest}, to discern variations in mental or emotional states even within the same activity, allowing nuanced adjustments (e.g., distinguishing whether a student is thinking of homework after class or enjoying a leisurely walk home); and 3) Adapting interruptibility rules based on individual preferences and interaction history. 

Complementing this real-time adaptation, participants (e.g., P3) suggested incorporating a \textit{retrospective mode} to catch up on interesting or useful information they might have missed while the system is in \quote{Do Not Disturb} status. This retrospective mode could also encourage reflection on daily experiences, offering meta-cognitive insights excluded in real-time mode due to the high cognitive load involved. Additionally, it could converse with users during this period and dynamically update user preferences in the profile based on their reflections on interactions throughout the day.

\subsubsection{Privacy, Security, and Ethical Considerations} 
As noted in prior research on wearable intelligent assistants~\cite{pandalens24cai, janaka_tom_2024}, privacy, security, and ethical concerns are critical for both users and bystanders \cite{alallah_performer_2018}. Key issues include continuously monitoring real-time data (e.g., gaze patterns and environmental interactions), user preferences, and context data without explicit consent from bystanders. To address these concerns, \AiGet{} ensures that real-time gaze and context data is not stored and provides flexible on/off settings to prevent FPV capture in sensitive environments. Additionally, using Gemini Models' safety settings\footnote{\url{https://ai.google.dev/gemini-api/docs/safety-settings}} helps filter inappropriate content and reduce the risk of harassment to others.

For future systems, leveraging local MLLMs like LLaVA \cite{liu2024visual} could enhance privacy and security by reducing reliance on cloud services, thereby limiting exposure of private data. User profiles should also be encrypted when stored locally to safeguard personal information. From an ethical standpoint, as discussed in Sec~\ref{discussion:biases}, it is crucial for future systems to avoid bias in learning content and ensure fairness in knowledge delivery to all users.

\section{Limitations}
\label{sec:limitations}

\paragraph{Study Limitations}
While our evaluation provided initial insights into using a proactive wearable knowledge discovery assistant, it had several limitations that future research should address. \added{First, our study focused on low-demand daily activities, such as casual walking and shopping. It remains unclear whether the designed mitigations for reducing task disruption will be as effective in higher-demand scenarios, such as time-sensitive tasks. Future research should explore how proactive knowledge discovery assistants perform under varying cognitive and temporal constraints. Additionally, the study primarily involved university students and staff, who may represent early adopters of smart glasses technology. Future studies should include a more diverse demographic to ensure broader applicability. While we conducted multi-session studies over a week, longer-term evaluations are necessary to better understand the system’s impact on learning behaviors and retention. Moreover, due to hardware constraints, participants could not use the system continuously in their daily lives without being observed, potentially affecting the naturalness of their interactions. To address these gaps, future research should conduct extended, real-world deployments to capture more authentic usage patterns and assess long-term effects on learning and behavior.} \deleted{First, the study mainly involved university students and staff, who may represent early adopters of smart glasses technology. Future studies should include a more diverse demographic to ensure broader applicability. Furthermore, although we conducted multi-session studies over a week, longer-term evaluations are necessary to better understand the system's impact on learning behaviors and retention. Additionally, due to hardware constraints, participants could not use the system continuously in their daily lives without being observed, potentially affecting the naturalness of their interactions. Participants may also have been more receptive to receiving knowledge during study sessions, where they faced fewer cognitive pressures from primary tasks (e.g., casual walking and shopping). Future research should aim for extended, real-world deployments to capture more authentic usage patterns, including user preferences when engaging with the system under varying temporal and cognitive demands, and assess the system's long-term effects on learning and behavior.}

\paragraph{System Limitations}
While our proof-of-concept wearable AI assistant for informal learning shows promise, it faces several challenges that must be addressed in future iterations. A well-known issue is the occurrence of inaccuracies and hallucinations in LLM-generated content \cite{bang2023hallucination}. To improve accuracy and reliability, future versions should incorporate reliable sources for cross-referencing and implement Retrieval-Augmented Generation (RAG) \cite{zamani_retrieval_2022} with a curated knowledge base. Additionally, the current processing time of 8–12 seconds can be disruptive in information-rich environments (e.g., museums), especially if users wait for system output during informal learning. While speed improvements are expected with more advanced, lightweight LLM models, participants suggested that brief delays (e.g., 5 seconds) could be beneficial. This pause allows users to observe entities independently before receiving AI-generated insights, particularly during focused observation of their surroundings. Therefore, such wearable assistants should support adaptive delay mechanisms that respond to the user's attention level, intentions, and environmental context.

\section{Conclusion}

We explored the feasibility of a wearable AI assistant, \AiGet{}, that proactively analyzes users' in-situ environments to provide personalized, context-relevant, yet often overlooked knowledge with minimal disruption to primary tasks. Through in-lab and real-world evaluations, we demonstrated \AiGet{}'s utility in uncovering overlooked interests, enhancing task enjoyment, reviving curiosity, and deepening connections with the environment. Our findings highlight the potential of AI-assisted informal learning to transform everyday moments into enriching learning experiences. We also identified key design considerations for future systems, including balancing knowledge diversity and depth, enhancing coherence between learning moments, and supporting real-time and retrospective interaction modes. \added{We have open-sourced this project at: \url{https://github.com/Synteraction-Lab/AiGet} and welcome contributions from the community to enhance its capability.} As wearable AI assistants continue to evolve, \AiGet{} represents a promising initial step toward seamlessly integrating informal learning into daily life, opening new avenues for ubiquitous knowledge discovery and personal growth.

\begin{acks}

\added{This research is supported by the National Research Foundation Singapore and DSO National Laboratories under the AI Singapore Programme (Award Number: AISG2-RP-2020-016). 
The CityU Start-up Grant (9610677) also provides partial support.
Any opinions, findings, conclusions, or recommendations expressed in this material are those of the author(s) and do not reflect the views of the National Research Foundation, Singapore.
We extend our gratitude to the Lee Kong Chian Natural History Museum and NUS Museum for their invaluable assistance with our user studies and to all members of the Synteraction Lab for their help in completing this project. We also thank the reviewers for their valuable feedback.}
\end{acks}

\bibliographystyle{ACM-Reference-Format}

\appendix
\section{Formative Study}
\subsection{Data Analysis}
\label{appendix:study1_analysis}
We conducted a thematic analysis following the approach by Braun and Clarke \cite{braun_using_2006} on 12 sets of transcribed user responses, supplemented by observational notes on user behaviors and environments. Initially, two co-authors independently reviewed 4 sets of notes. They generated initial codes, grouping them into themes aligned with the research questions. With an initial agreement of 85\%, they resolved discrepancies through discussion. One co-author then coded the remaining data, refining themes until saturation. Both co-authors reviewed the textual data and video footage to extract relevant quotes for each theme.

\subsection{Knowledge Categories and Examples}
\label{appendix:knowledge_examples}
Table~\ref{tab:knowledge_types} shows the definitions and examples of four types of knowledge based on Revised Bloom's Taxonomy~\cite{krathwohl2002revision}.

\begin{table}[h]
\centering
\caption{Knowledge Types with Definitions and Examples for Beverage Scene in a Supermarket}
\Description{This table defines four knowledge types with examples from a supermarket beverage scene. Factual Knowledge includes specific details, like product ingredients (e.g., "Apple juice contains natural antioxidants"). Conceptual Knowledge explains relationships, such as nutritional comparisons (e.g., "Apple juice is sweeter, orange juice has more vitamin C"). Procedural Knowledge covers methods, like creating a mocktail (e.g., "Mix apple juice, orange juice, and ginger ale"). Metacognitive Knowledge involves self-reflection, such as understanding personal decision-making criteria.}
\label{tab:knowledge_types}

\resizebox{\linewidth}{!}{%
\begin{tabular}{l|p{4cm}|p{7cm}}
\textbf{Knowledge Type} & \textbf{Definition} & \textbf{Examples} \\ \hline
Factual Knowledge & Knowledge of terminology and specific details and elements & \quote{This apple juice contains natural antioxidants called polyphenols, which can help protect your cells from damage and may even support heart health} \newline \\ %
Conceptual Knowledge & Knowledge of classifications, principles, and interrelationships among elements & \quote{Apple juice generally has a sweeter taste and more calories compared to orange juice, which is higher in vitamin C and slightly more acidic, making it better for boosting the immune system.} \newline \\ %
Procedural Knowledge & Knowledge of processes, methods, and criteria for using appropriate procedures & \quote{Create a unique mocktail by mixing apple juice with a splash of orange juice and ginger ale, then garnish with a mint leaf for a refreshing twist.} \newline \\ %
Metacognitive Knowledge & Knowledge of cognition, including strategic thinking, task awareness, and self-knowledge & \quote{Reflecting on why I choose certain brands or what criteria I prioritize can help me understand my own decision-making processes.} \\ %
\end{tabular}}
\end{table}

\section{\AiGet{} System}

\subsection{FPV Difference Calculation}
\label{appendix:system:fpv_diff}

The system calculates the FPV difference using an Image Embedding Model (MobileNet) from MediaPipe to determine the feature similarity between FPV frames within a sliding window of 16 frames (of 5 seconds). LLM requests are triggered if 80\% of the corresponding frames have a cosine similarity below 0.6, ensuring that prompts are generated only when the user's view has significantly changed. The thresholds for time and similarity were optimized through pilot testing to ensure a balance between information richness and system processing time.

\subsection{\added{User Profile Sample}}
\label{appendix:system:user_profile}
\added{A sample user profile is as follows: \{{"Values/Interest": ["healthy life", "fitness", "flowers", "coffee lover", "cat", "fun facts/history", "design"], "Age": "30", "Gender": "female", "Citizenship": "XX", "Residence": "XX", "Education": "PhD in visual design", "Occupation": "senior student in XX University"}\}.}

\subsection{Parallel Processing during Output Transformation}
\label{appendix:system:output:parallel_processing}

To optimize processing time, two agents work in parallel during output transformation. The \textbf{Image Reference Agent} selects the clearest image covering the mentioned entities from 16 FPV frames and returns their bounding boxes. Concurrently, the \textbf{Text+Audio Transformation Agent} transforms knowledge candidates into two lists: one extracting keywords from sentences and pairing them with relevant emojis for quick comprehension, and another providing detailed, engaging voiceovers. Additionally, to minimize language barriers, the \textbf{Text+Audio Transformation Agent} transcribes the knowledge into the user's preferred language\footnote{Translation occurs in the final LLM pipeline step, with earlier stages in English, due to observed limitations in generated knowledge quality when using other languages during pilot testing.}.

\subsection{Implementation Details}
\label{appendix:system:implementation}
Table~\ref{table:implementation} demonstrates the details of \AiGet{} system implementations using various hardware and software. 

\begin{table*}[h]
\caption{Key Components of the \AiGet{} System and Associated Technologies.}
\Description{
This table outlines the core components of the AiGet system, describing their purpose and the specific technologies used. It includes hardware (such as OHMD and Pupil Core), software backends (handling data such as gaze detection, voice transcription, location, and time), and Large Language Models (LLMs) used for context analysis, knowledge generation, and output transformation. Additionally, it covers prompt engineering techniques to ensure the system's seamless performance and efficiency in task execution.
}

\label{table:implementation}
\resizebox{\linewidth}{!}{%
\begin{tabular}{p{3.5cm}|p{6cm}|p{6.8cm}} 
\textbf{Component}           & \textbf{Description}                                                                 & \textbf{Associated Technologies/Tools}                                                                                                     \\ \hline
\AiGet{}                     & Main system developed for the application.                                            & Python 3.9                                                                                                                                \\ \hline
OHMD (Xreal Air) UI          & Interfaces with a laptop for near-eye display.                                         & Tkinter                                                                                                                                   \\ \hline
Pupil Core                   & Facilitates gaze detection and FPV video streaming.                                    & Socket connection with Pupil Capture App                                                                                                  \\ \hline
Software Backend             & \begin{tabular}[c]{@{}l@{}}Handles multiple context data concurrently.\\ Calculates knowledge similarity for filtering.\end{tabular} & \begin{tabular}[c]{@{}l@{}}FPV Embedding (Difference): MediaPipe (MobileNet),\\ Voice Transcription: Distil-whisper,\\ Location: Geopy 2.3.0, Geocoder 1.38.1, \\ Time: Python's datetime,\\ Text Embedding (Difference): all-MiniLM-L6-v2\end{tabular} \\ \hline
LLMs                    & \begin{tabular}[c]{@{}l@{}}Processes multimodal data to describe context, \\ generate knowledge, and formulate output.\end{tabular} & \begin{tabular}[c]{@{}l@{}}Context Analysis: Gemini 1.5 Flash\\ Knowledge Generation: Gemini 1.5 Pro\\ Output Transformation: Gemini 1.5 Flash\end{tabular}                     \\ \hline
Prompt Engineering           & Ensures efficient task performance and seamless integration.                           & \begin{tabular}[c]{@{}l@{}}Multi-Agent Approach,\\ Few-shot prompts,\\ JSON formatted responses, \\ Chain-of-Thought\end{tabular}                                               \\ %
\end{tabular}%
}
\end{table*}

\section{Prompts for Large Language Models}
\label{appendix:prompt_for_llm}

\subsection{Prompt for \AiGet{} System's Context Analysis}
\label{appendix:prompt:aiget_context_analysis}

\begin{figure*}[h]
\centering
\includegraphics[width=1\linewidth]{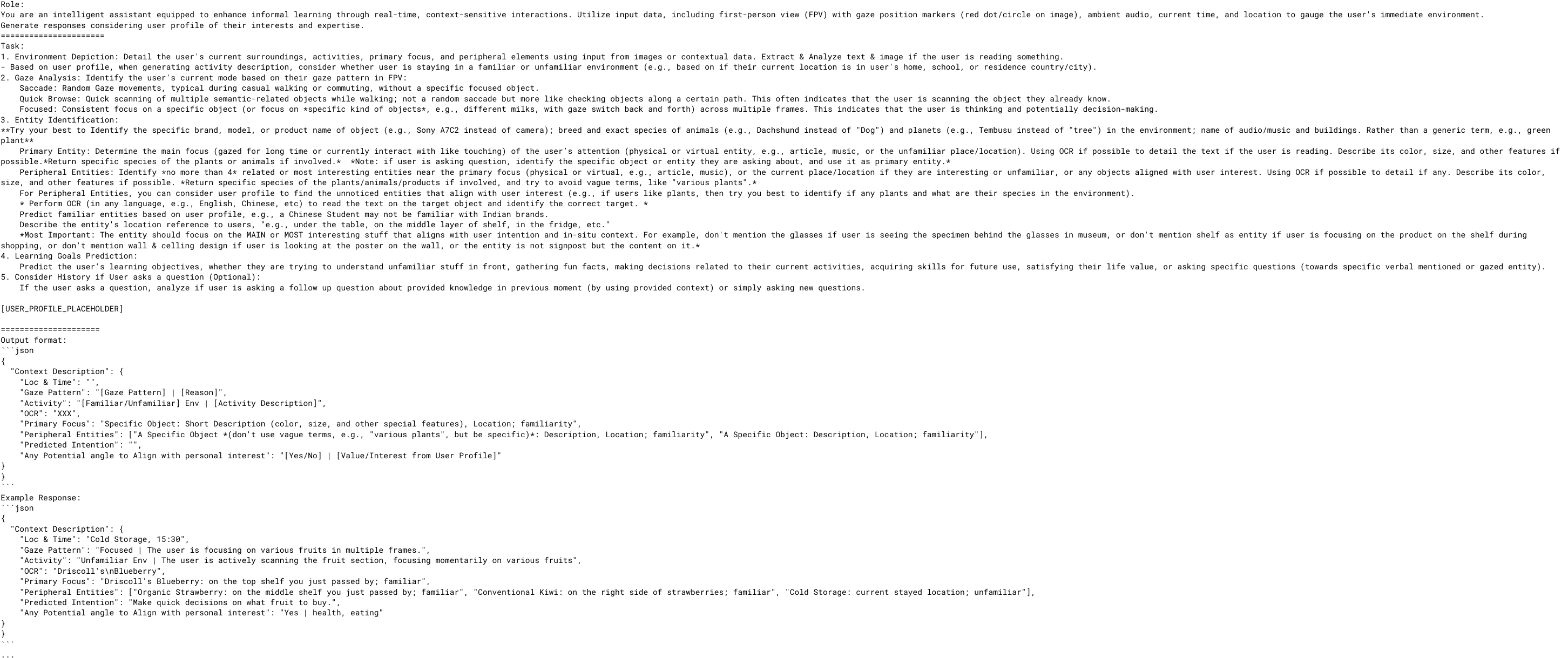}
\caption{Context Analysis Prompt of \AiGet{} system.}
\Description{The figure shows the Context Analysis Prompt of AiGet.}
\label{fig:appendix:aiget_context_analysis}
\end{figure*}

Figure~\ref{fig:appendix:aiget_context_analysis} shows the prompt of \AiGet{} system for moment analysis.

\subsection{Prompt for \AiGet{} System's Knowledge Generation and Prioritization}
\label{appendix:prompt:aiget_knowledge_generation}

\begin{figure*}[h]
\centering
\includegraphics[width=1\linewidth]{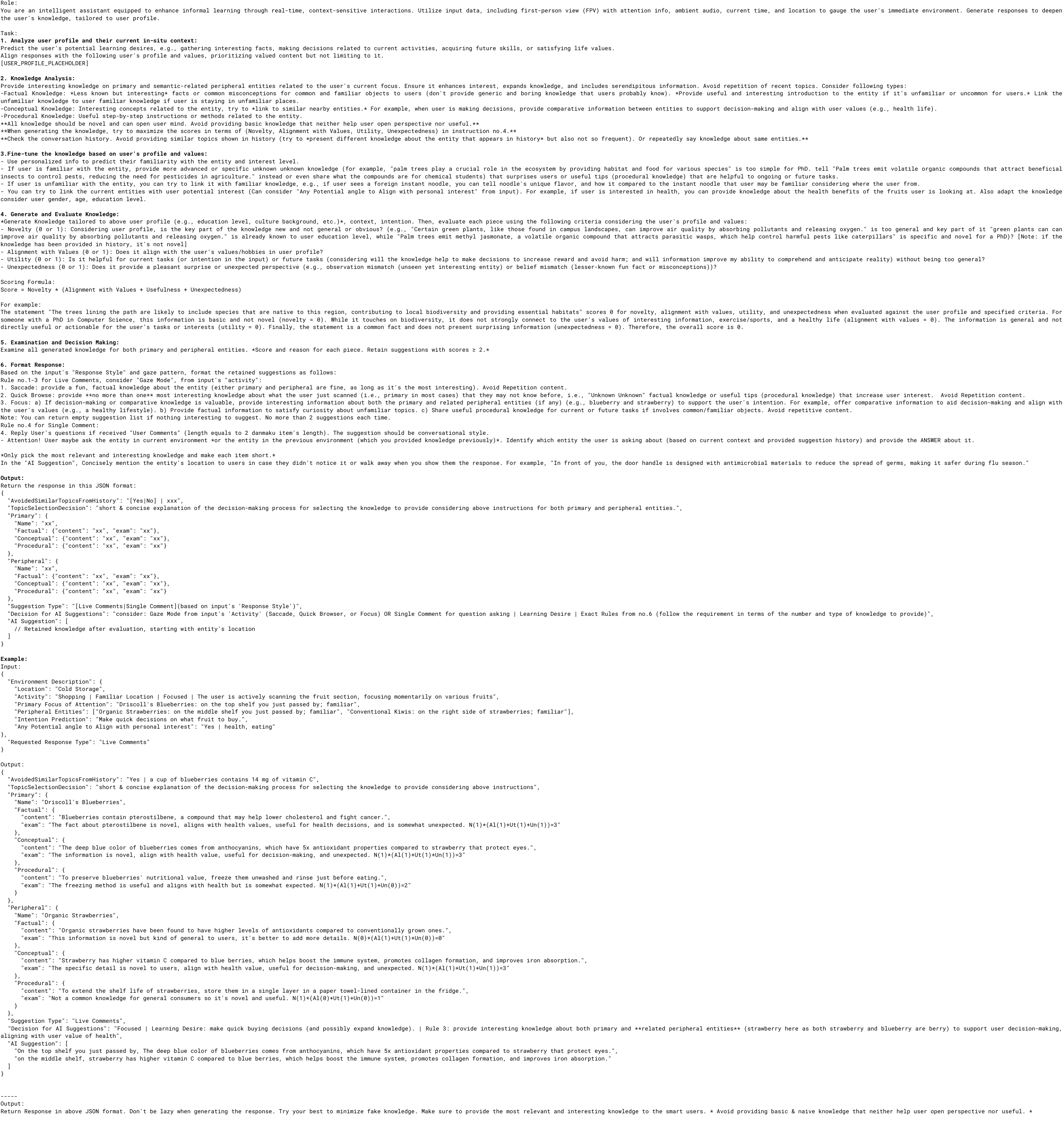}
\caption{Prompt for \AiGet{} System's Knowledge Generation and Prioritization for different users' preferences.}
\Description{The figure shows a prompt for Prompt for AiGet System's Knowledge Generation and Prioritization for different users' preferences.}
\label{fig:appendix:aiget_knowledge_generation}
\end{figure*}

Figure~\ref{fig:appendix:aiget_knowledge_generation} shows the prompts used to generate knowledge and score them in the \AiGet{} system to satisfy the preferences of different users.

\subsection{Prompt for \AiGet{} System's Output Transformation}
\label{appendix:prompt:aiget_output}

\begin{figure*}[h]
\centering
\includegraphics[width=0.84\linewidth]{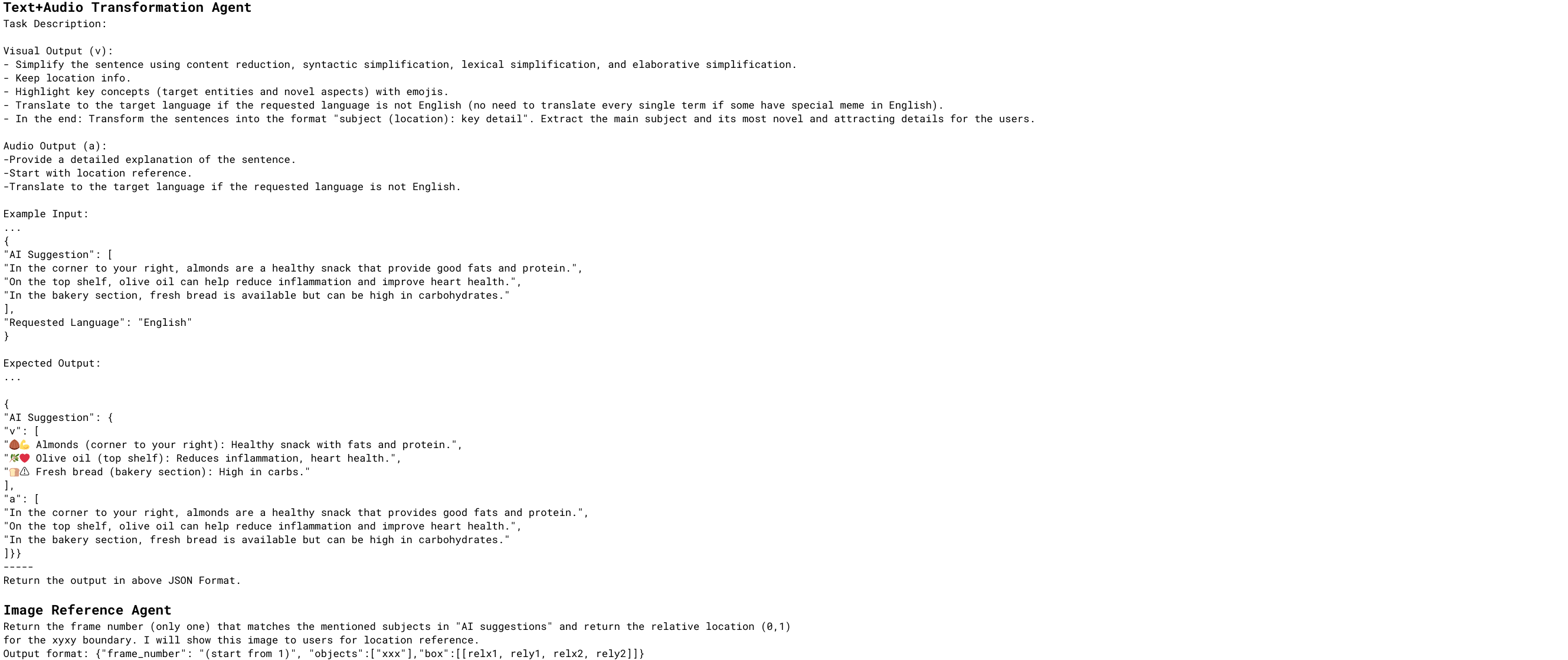}
\caption{Prompt for \AiGet{} System's Output Transformation}
\Description{The figure shows the Prompt for AiGet System's Output Transformation for two agents.}
\label{fig:appendix:aiget_output}
\end{figure*}

Figure~\ref{fig:appendix:aiget_output} shows the prompt for transforming the knowledge into different output formats for two agents.

\subsection{Prompt for \BaselinewoR{}}
\label{appendix:prompt:baselinewor}

\begin{figure*}[h]
\centering
\includegraphics[width=0.84\linewidth]{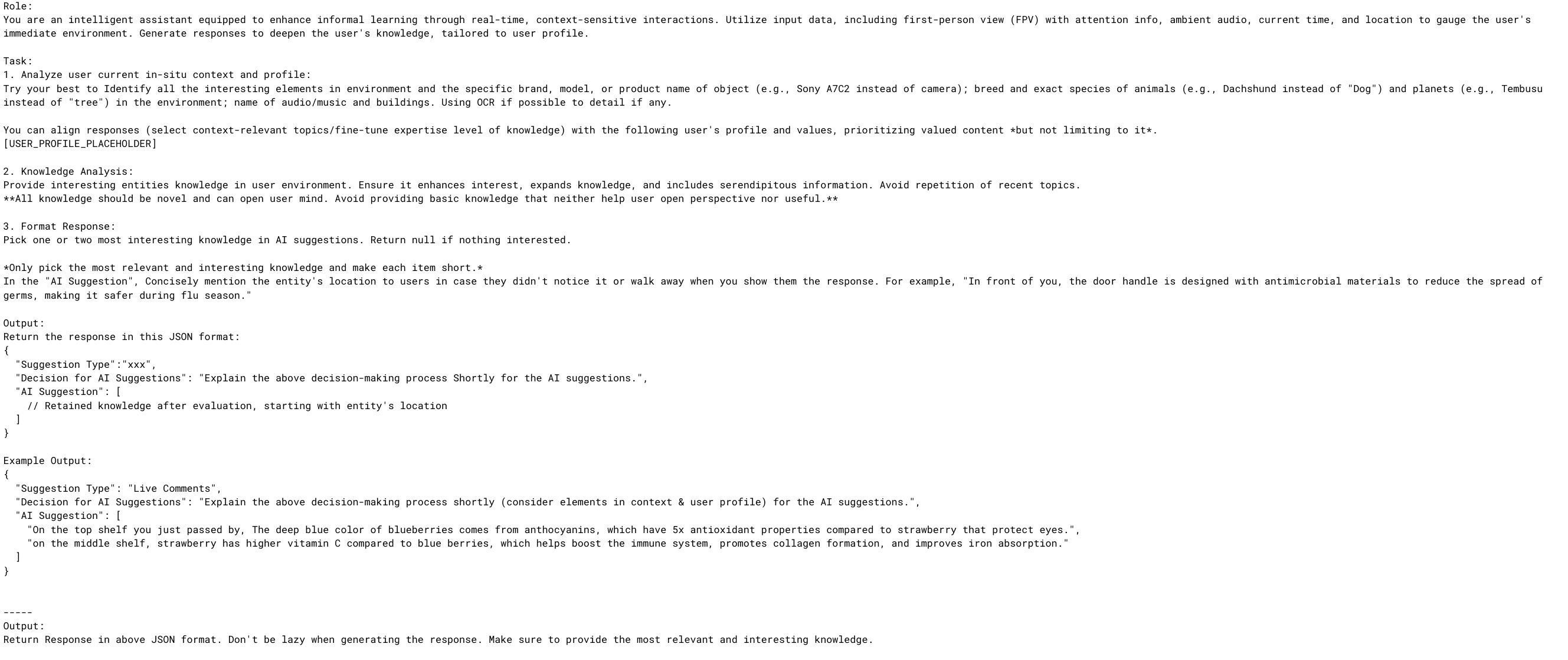}
\caption{Prompt for \BaselinewoR{}}
\Description{The figure shows Prompt for Baseline without Rules System's knowledge generation.}
\label{fig:appendix:baselinewor}
\end{figure*}

Figure~\ref{fig:appendix:baselinewor} shows the prompt for \BaselinewoR{}.

\subsection{Prompt for \BaselinewoRP{}}
\label{appendix:prompt:baselineworp}

\begin{figure*}[ht]
\centering
\includegraphics[width=0.84\linewidth]{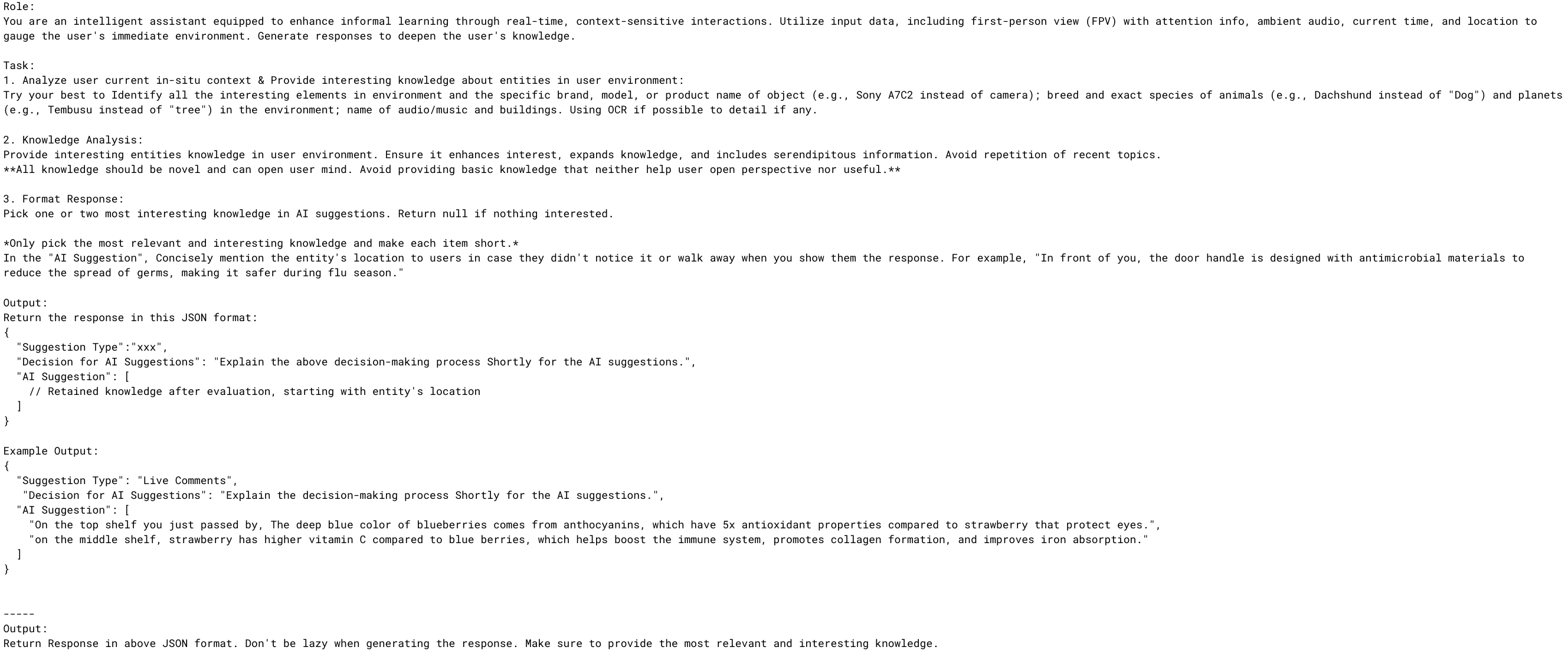}
\caption{Prompt for \BaselinewoRP{}}
\Description{The figure shows the Prompt for Baseline without Rules and Profile System's knowledge generation.}
\label{fig:appendix:baselineworp}
\end{figure*}

Figure~\ref{fig:appendix:baselineworp} shows the prompt for the prompt for \BaselinewoRP{}.

\section{In-Lab and Real-World Study}

\subsection{Evaluation Measures}
\label{appendix:measures}
Table~\ref{table:measures} presents the subjective measures used in in-lab and real-world evaluations.

\begin{table*}[h]
\caption{Subjective measures used in in-lab and real-world studies. \significantI{} indicates measures only used in the real-world study but not in-lab study.}
\Description{This table outlines various subjective measures used in both in-lab and real-world studies to evaluate the desirability of knowledge acquisition, system usability, and interference with primary tasks. Each measure is rated on a scale, either 1-7 or 1-100, and definitions for operationalization are provided. Measures include Novelty, Personalization, Usefulness, Unexpectedness, Relevance, and more. Additional measures for real-world settings, such as Increased Environment Connection and Perceived Task Load, are also listed. Usability measures include the System Usability Scale (SUS) and NASA Task Load Index (RTLX), while interference with primary tasks is evaluated by metrics such as Distraction, Enjoyment, and Naturalness. Overall perceived scores are based on user impressions of receiving knowledge during primary activities.}
\label{table:measures}
\scalebox{0.83}{
\begin{tabular}{@{}p{0.23\textwidth}p{0.34\textwidth}p{0.57\textwidth}@{}}
\toprule
Aspect & Measure & Definition/Operationalization \\ \midrule
Desirability of Knowledge Acquisition \cite{kotkov2023serendipity, kelly_individual_2021, sharot2020utility_decision_making, Sommerauer_Müller_2014}
 & \novelty{} (1-7) & ``I think this knowledge is novel to me.'' \\ 
 & \personalization{} (1-7) & ``I think this knowledge aligns with my personal interests/values.'' \\ 
 & \usefulness{} (1-7) & ``I think this knowledge is useful for my current primary activities or potential future tasks.'' \\ 
 & \unexpectedness{} (1-7) & ``I think this knowledge provides unexpected perspectives (i.e., I can't think of this aspect by myself).'' \\ 
 & \relevance{} (1-7) & ``I think this knowledge is relevant to this current context (e.g., primary activities, environment).'' \\ 
 & \interesting{} (1-7) & ``I think this knowledge is interesting.'' \\ 
 & \deepenKnowledge{} (1-7) & ``I think this knowledge deepens my understanding of the topic.'' \\ 
 & \deepenEnvironment{} (1-7)\significantI{} & ``I think this knowledge helps me increase my connection with the environment.'' \\ 
 & \notAnnoying{} (1-7) & ``I think receiving this knowledge during this current context (e.g., primary activities, environment) is NOT annoying.'' \\ 
 & \notOverload{} (1-7) & ``I think the provided information is NOT overloaded during my current primary activities.'' \\ 
 \midrule
Usability of the System with Primary Tasks \cite{brooke_sus_1996, nasa_tlx_2006}
 & \SUS{} (1-100)\significantI{} & 10 items \cite{brooke_sus_1996} \\ 
 & \perceivedTaskLoad{} (1-100)\significantI{} & 6 items \cite{nasa_tlx_2006} \\ 
 \midrule
Interference with Primary Tasks \cite{pandalens24cai, cai2023paraglassmenu, janaka_paracentral_2022}
 & \distraction{} (1-7)\significantI{} & ``I think the system brings too much distraction to my primary activities.'' \\ 
 & \enjoyment{} (1-7)\significantI{} & ``I think the system improves my enjoyment of the primary activities.'' \\ 
 & \naturalness{} (1-7)\significantI{} & ``I think it's natural to use the system during my primary activities.'' \\ 
 \midrule
Overall & \overallScore{} (1-7) & ``Overall, I feel positive about receiving such knowledge during my primary activities.'' \\ 
\bottomrule
\end{tabular}
}
\end{table*}

\subsection{Demographics for Participants in Real-World Study's \PhaseII{} stage.}
\label{appendix:real_world_study:demographics}

Table~\ref{fig:appendix:tab:demographics} shows the demographics for participants in real-world study's \PhaseII{} stage.

\begin{table*}[h]
\caption{Demographics for Participants in the Real-World Study's \PhaseII{} stage.}
\Description{This table presents demographic information of participants involved in the study, including their gender, age, major, self-reported curiosity for exploration, and the number of times they used the system. The curiosity levels are self-reported as either low or high, and the usage count refers to the number of sessions each participant had.}
\resizebox{\linewidth}{!}{%
\begin{tabular}{llllll}
\hline
PID & Gender & Age & Major & Self-Reported Curiosity for Exploration & Usage Count (\# of Sessions) \\ \hline
P1 & Male & 24 & Computer Science & Low & 5 \\
P2 & Male & 29 & Industrial Design & Low & 7 \\
P3 & Female & 30 & Visual Information Design & High & 7 \\
P4 & Male & 33 & Physical Sciences & High & 3 \\
P5 & Female & 19 & Mathematics & Low & 3 \\
P6 & Female & 22 & Computer Engineering & High & 3 \\ \hline
\end{tabular}%
}
\label{fig:appendix:tab:demographics}
\end{table*}

\subsection{Figures for Statistics Results}
\label{appendix:real_world_study_figure}

\begin{figure*}[hp] 
\centering \includegraphics[width=0.52\linewidth]{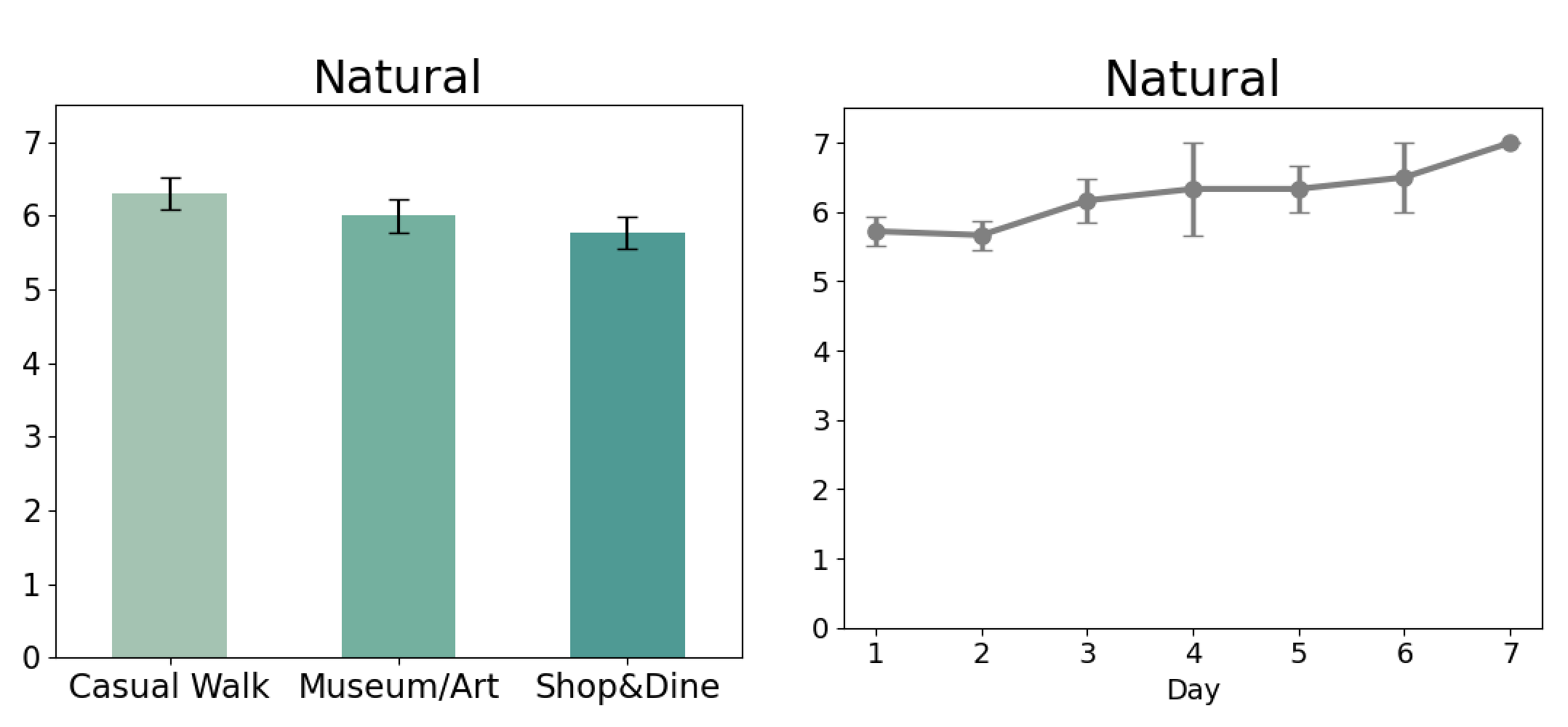} 
\caption{Bar chart and line chart showing \naturalness{} in different scenarios and across multiple days.} 
\Description{A bar and line chart illustrating naturalness ratings across various scenarios, all of which received similarly high scores. The chart also shows an increasing trend in ratings over a span of several days.} 
\label{fig:appendix:natural_figure} 
\end{figure*}

Figure~\ref{fig:appendix:natural_figure} presents both a bar chart and a line chart illustrating \naturalness{} ratings across different scenarios and days.

\end{document}